\pdfoutput=1
\RequirePackage{fix-cm}
\documentclass[smallextended]{svjour3}       
\smartqed  
\usepackage[numbers]{natbib}

\usepackage{dblfloatfix}
\usepackage{arydshln}

\usepackage{balance}
\usepackage{alltt}
\usepackage{subcaption}

\usepackage{pifont}
\usepackage{graphicx}
\usepackage{colortbl}
\usepackage[dvipsnames]{xcolor}
\usepackage{rotating}
\usepackage{makecell}
\usepackage{tabularx}
\usepackage{booktabs}
\usepackage{wrapfig}
\usepackage{tikz}
\usetikzlibrary{angles}
\usepackage{multirow}
\usepackage{hyperref}

\usepackage{framed}
\usepackage[framemethod=tikz]{mdframed}
\usetikzlibrary{shadows}
\usepackage{enumitem}
\usepackage{graphics}
\usepackage[para,online,flushleft]{threeparttable}

\setlist[description]{leftmargin=1cm}

\usepackage{amsmath}

\usepackage{pgfplots}
\usetikzlibrary{patterns}
\usetikzlibrary{arrows}
\usetikzlibrary {positioning}
\usepackage{enumitem}
\setlist[itemize]{leftmargin=*}
\setlist[enumerate]{leftmargin=*}
\usepackage{makecell}
\usepackage{xcolor,pifont}

\usepackage[linesnumbered,ruled,vlined]{algorithm2e}
\setlist{nolistsep} 

\SetKwProg{Fn}{Function}{}{}

\setlist[1]{itemsep=0pt}

\newmdenv[
tikzsetting= {fill=gray!10},
linewidth=1pt,
roundcorner=2pt,
shadow=false
]{myshadowbox}

\newcolumntype{P}[1]{>{\centering\arraybackslash}p{#1}}

\hypersetup{
    linkcolor=blue,
    filecolor=magenta,      
    urlcolor=cyan,
}

\newcommand{\bi}{\begin{itemize}[leftmargin=0.4cm]}
\newcommand{\ei}{\end{itemize}}
\newcommand{\be}{\begin{enumerate}[leftmargin=0.4cm]}
\newcommand{\ee}{\end{enumerate}}

\usepackage{url}

\tikzstyle{thmbox} = [rectangle, rounded corners, draw=black, fill=gray!10]

\usepackage[tikz]{bclogo}
\usepackage{transparent}

\definecolor{cyan}{HTML}{2FF3E0}
\newenvironment{RQ}[1]%
{\noindent\begin{minipage}[c]{\linewidth}%
\begin{bclogo}[couleur=gray!30,%
                arrondi=0.1,%
                logo=\bctrombone,%
                ombre=true]{{\normalsize ~#1}}}%
{\end{bclogo}\end{minipage}\vspace{2mm}}

\newcommand*\colourcheck[1]{%
  \expandafter\newcommand\csname #1check\endcsname{\textcolor{#1}{\ding{51}}}%
}

\colourcheck{green}

\newcommand*\colourx[1]{%
  \expandafter\newcommand\csname #1x\endcsname{\textcolor{#1}{\ding{53}}}%
}
\colourx{red}

\usepackage{listings}
\usepackage{cellspace}
\newcommand{\IT}{\textbf{DebtFree}}

\newcommand{\revised}{\textcolor{blue}}
\newcommand{\respto}[1]{
\fcolorbox{black}{black!15}{
\label{response:#1}
\bf
  \scriptsize R-{#1}}~
}

\begin{document}

\title{\textbf{DebtFree}: Minimizing Labeling Cost in Self-Admitted Technical Debt Identification using Semi-Supervised Learning}
\titlerunning{DebtFree: Minimizing Labeling Cost in SATD Identification using SSL}

\author{Huy Tu         \and
         Tim Menzies
}

\institute{H. Tu and T. Menzies \at
              Department of Computer Science, \\
              North Carolina State University, \\
              Raleigh, USA \\
              \email{hqtu@ncsu.edu} and timm@ieee.org}





\date{Received: date / Accepted: date}

\maketitle

\begin{abstract}

Keeping track of and managing Self-Admitted Technical Debts (SATDs) is important for maintaining a healthy software project. 
Current active-learning SATD
recognition tool involves manual inspection of 24\% of the test comments on average to reach 90\% of the recall. Among all the test comments, about 5\% are SATDs. The human experts are then required to read almost a quintuple of the SATD comments which indicates the inefficiency of the tool. Plus, human experts are still prone to error: 95\% of the false-positive labels from previous work were actually true positives. 


To solve the above problems, we propose DebtFree, a two-mode framework based on unsupervised learning for identifying SATDs. 
In mode1, when the existing training data is unlabeled, DebtFree starts with an unsupervised learner to automatically pseudo-label the programming comments in the training data.
In contrasts, in mode2 where labels are available with the corresponding training data,  DebtFree starts with
a pre-processor that identifies the highly prone SATDs from the test dataset.
Then, our machine learning model is employed to assist human experts in manually identifying the remaining SATDs. Our experiments on 10 software projects show that both models yield statistically significant improvement in effectiveness over the state-of-the-art automated and semi-automated models. Specifically, DebtFree can reduce the labeling effort by 99\% in mode1 (unlabeled training data), and up to $63\%$ in mode2 (labeled training data) while improving the current active learner's F1 relatively to almost 100\%.




\end{abstract}



\section{Introduction}\label{sec:introduction}
\vspace{-8pt}


When developers  rush out code, that code
often contains {\em technical debt} (TD), i.e. decisions that must later be repaid with further work.
As the initial step towards understanding and resolving TDs, many research~\cite{guo2019mat, ren2019neural, jitterbug} first detecting the intentionally documented (via comments) TD, i.e., self-admitted TD (SATD). SATDs are crucial to identify as they (1) are diffused
in the codebase~\cite{jitterbug}; (2) can survive long-term
(more than 1,000 commits)~\cite{bavota2016large}; and (3) complicate maintainability of the software~\cite{Fucci_MSR_21, wehaibi2016examining}
State-of-the-art (SOTA) works
have identified SATDs automatically \cite{ren2019neural} or semi-automatically 
\cite{jitterbug}.

However, models that recognize
TD must be learned from  {\em
labeled data}. Generating such labels can be extremely slow and expensive.
For instance, 
Tu et al.~\cite{tu2020better} reported that manually labeling $22,500+$ commits 
required 175 person-hours, including cross-checking.
Due to the labor-intensive nature of the process, researchers often reuse
datasets labeled from previous studies. For instance, Lo et al., Yang et al., and Xia et al. certified their methods using data generated by Kamei et al. \cite{kamei12_jit,xia16ist17,yang2015deep,yang16unsupervised}. While this practice allows researchers to rapidly
test new methods, it leaves the possibility for  
any labeling mistake
to propagate to other related works. In fact, before reusing Maldonaldo et al.'s data~\cite{maldonado2015detecting} to identify SATDs,  Yu et al.~\cite{jitterbug} discovered that more than 98\% of the false positives were actually true positives, casting doubt on related work using the original dataset.
Hence, it is timely to ask:
\begin{quote}
\centering
\vspace{-8pt}
{\em
\hspace{-10pt}Can we reduce the labeling effort associated
with building models for  technical
debt?
}
\vspace{-8pt}
\end{quote}
An unsupervised learning technique that learns patterns from unlabeled data is a promising direction in SATD identification. However, without supervision, the technique alone can be ineffective. As illustrated in 
Figure \ref{fig:workflow}, our approach is to first demonstrate that previous methods can be extended or integrated with unsupervised learning to greatly reduce the labeling effort while effectively recognizing SATDs. This proposed method, called \IT{},  includes the combination of three separate
approaches in a novel manner:  
 
\be
\item \textbf{Pseudo-Labeling}: This step is required if the training data does not have any labels to start with. First, we frugally pseudo-labels the training data with unsupervised learning, i.e., identifying hidden patterns in data in order to map unlabeled examples in two groups. Intuitively, the more complex the data instance~\cite{tu2021frugal}, the more likely that the comment is describing a SATD.
CLA by Nam et al.\cite{nam2015clami} is an example of an unsupervised classifier that recognizes ``complex'' examples (those with many values above the median). The intuition here is well documented~\cite{menzies07defect, tu2021frugal, DAmbros2012, hassan_icse09}.

\item \textbf{Filtering}: This step is optional. We identify early and remove instances from the test dataset that are likely to be SATDs.

 \item \textbf{Active Learning}: This step is always required. We train on some labeled data and then guide the human experts to manually find the comments that are most likely to contain SATDs. It is critical to assess whether the labeled data is insightful enough to guide the human experts for the entire labeling process. 
 If not, 
 we propose Falcon, a new active learning policy to take advantage of such data while still ensuring effectiveness. 

\ee In this work, we aim to better data generation associated with building models for SATDs identification by reducing the labeling effort come from manual method~\cite{tu2020better} and improving the labeling quality of fully automated methods~\cite{jitterbug}. Moreover, our investigation also showed that the effort-aware method we propose, \IT{}, also performs statistically similar or even better than two SOTA works \cite{jitterbug, ren2019neural}. To understand and validate this end-to-end method, \IT{}, we investigate the following research questions:

\begin{figure*}[!tbp]
\vspace{-7pt}
\begin{center}

\includegraphics[width=\textwidth]{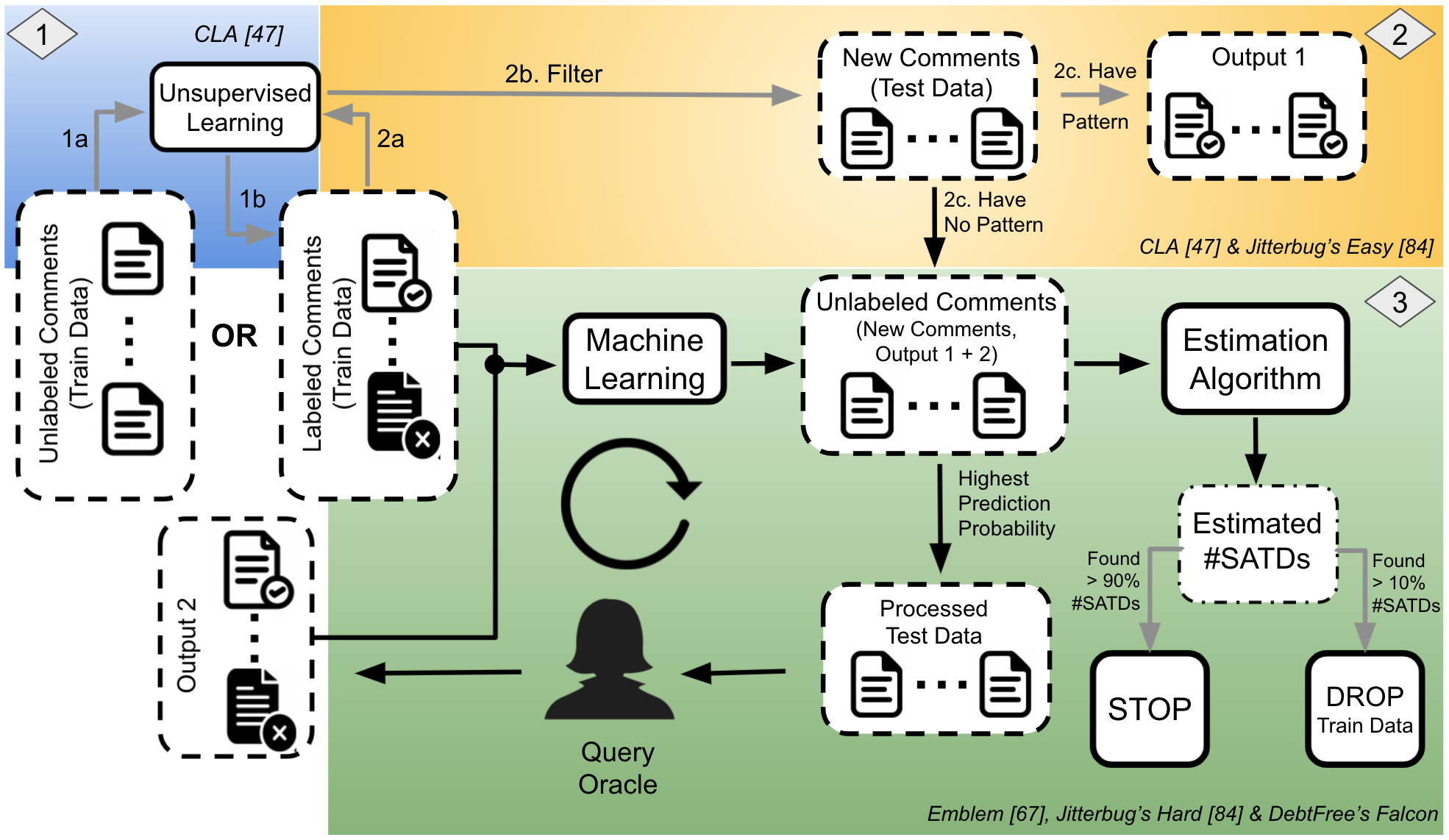}
\vspace{-18pt}
\caption{Workflows of \IT{} = Pseudo-Labeling (via Unsupervised Learning, i.e., CLA~\cite{nam2015clami}) + Filtering (via CLA~\cite{nam2015clami} or Jitterbug's Easy~\cite{jitterbug}) + Active Learning (via Emblem~\cite{tu2020better}, Jitterbug's Hard~\cite{jitterbug}, or this study's Falcon). Step 1 is required if there is no access to the training data's labels. Step 2 is optional while Step 3 is required at all times. The gray arrows indicate different configurations of the method that will be investigated for this study.}
\label{fig:workflow}

\end{center}
\vspace{-30pt}
\end{figure*}

\textbf{RQ1: How well can the state-of-the-art unsupervised learning method identify SATDs?} We investigate variants that stem from Nam et al.'s unsupervised learning CLA method. As these methods leverage on hidden patterns within the data, we compare them to the pattern-based SOTA for identifying SATDs, also by~\citet{jitterbug}. \vspace{-10pt}

\begin{RQ}{\normalsize{Result:}} 
From our exploration of various unsupervised learners, the original CLA by Nam et al. performs the best. Moreover, CLA performs similarly to the SOTA pattern-based approach, Easy~\cite{jitterbug}, without having access to the data's labels (100\% less effort).
\end{RQ} \vspace{-15pt}

\textbf{RQ2: How can the state-of-the-art active learning framework be combined with the state-of-the-art unsupervised learner?} From RQ1, unsupervised learning methods are promising but not optimal. Hence, we study different combinations by incorporating a chosen unsupervised learner with several SOTA active learning frameworks in SE. \vspace{-11pt}

\begin{RQ}{\vspace{-30pt}\normalsize{Result:}} \vspace{-2pt}
With the effort-aware theme, we investigate different combinations of active learners and CLA across two settings of the training data, either with having 1-\textit{no access} and 2-\textit{access} to the labels to propose \IT{}. In setting 1, or \IT{}(0), the best combination is Pseudo-Labeling (via CLA) with our proposed active learner, Falcon. In setting 2, or \IT{}(100), the best combination is Filtering (via CLA) with the SOTA active learner for SATDs identification, Hard.
\end{RQ} \vspace{-16pt}

\textbf{RQ3: How does the proposed \IT{} perform against state-of-the-art models in identifying SATDs?} After finalizing two combinations from RQ2 to propose \IT{}(0)/(100), it is essential to assess their usefulness by comparing against the SOTA models for SATDs identification. \vspace{-11pt}

\begin{RQ}{\normalsize{Result:}} \vspace{-2pt}
When comparing  against the SOTA semi-supervised learning work by \citet{jitterbug} and the SOTA supervised learning work (with deep learning) by \citet{ren2019neural}, our proposed method \IT{} outperforms them significantly. First, \IT{}(100) performs similarly to \citet{ren2019neural}'s work and better than \citet{jitterbug}'s work while reducing the labeling cost by 2.5 times. Second, \IT{}(0) performs similarly to \citet{ren2019neural}'s work without having access to the training data's labels and outperforms \citet{jitterbug}'s work while expending 99\% less effort.
\end{RQ} \vspace{-16pt}

Our contributions to the field of software analytic are: 
\be
\item This work is the first to assess the usage of unsupervised learning to reduce the cost of labels labeling in identifying SATDs. 
\item In the low-resource setting (training data with no label), our unsupervised methods outperform the prior SOTA models while requiring less knowledge (prior work used 100\% labeled data while we get by with very little)~\cite{ren2019neural, jitterbug}. Counting the training data, we can reduce $99\%$ of the number of examples that have to be labeled. This is the largest reduction ever reported  in the effort required to commission a SATD identification model. 
\item In the high-resource setting (training data with labels), we propose an improvement to the two-step Jitterbug technique~\cite{jitterbug} by replacing the pattern-based approach with an unsupervised learner to help reduce the commissioning effort of labeling on new data by $62.5\%$ ($5/8$). 
\item Our proposed active learning scheme, Falcon, outperforms both SOTA deep learning method~\cite{ren2019neural} and SOTA two-step method~\cite{jitterbug} across both low-resource  and high-resource settings.

\item
Nearly all the prior work on unsupervised learning 
focus on defect prediction~\cite{fu2017unsupervised,nam2015clami,9115238,unsup_review,yan2017file,yang2016defect,yang16effort,yang16unsupervised,zhang2016cross,Zhou2018HowFW}. The performance of our framework suggests that many more domains in software analytics could benefit from unsupervised learning. 

   

\item To better support other researchers our scripts and data are on-line at https://github.com/HuyTu7/DebtFree.

\ee
The rest of this paper is structured as follows. \textit{Section 2} discusses the motivation, background and related works. \textit{Section 3} describes our methodology. \textit{Section 4} focuses on our experimental design, while \textit{section 5} analyzes the results. \textit{Section 6} and \textit{7} discuss our short-comings and directions for future work, respectively.



\section{Motivation and Background} \label{background}

\subsection{On the merits of studying Technical Debt and SATDs} \label{super_learning}

Technical Debts (TDs) are introduced in the software when developers make decisions based on short-term benefits instead of long-term stability. TDs
can accumulate interest similar to financial debts if they are not resolved in a timely manner. In 2012, after interviewing 35 software developers from diverse projects in different companies, varying both in size and type,~\citet{lim2012balancing}
found developers generate technical debts due to factors like increased workload,  unrealistic deadline in projects, lack of knowledge, boredom,
 peer-pressure among developers, unawareness or short-term business goals of stakeholders, and reuse of legacy, third-party, or open-source code. After observing five large-scale projects and companies in two studies,~\citet{wehaibi2016examining} and~\citet{martini2015danger} found that the number of technical debts in a project may be very low (only 3\% on average). However, those TDs contaminate other parts of a software system and create a significant amount of defects in the future. Fixing such technical debts is more difficult than regular defects, often twice the cost if not resolving immediately~\cite{guo2011tracking}). The Software Improvement Group study by~\citet{nugroho2011empirical} offers a cost estimate of TD accumulation:  a regular mid-level project owes $\$857,500$ in TD and resolving TD has a Return On Investment of 15\% in seven years.  Yet, limited success has been achieved despite a large body of research on identifying TD as part of Code Smells using static code analysis~\cite{fontana2012investigating,marinescu2004detection,marinescu2012assessing,marinescu2010incode,zazworka2013case}. Static code analysis has a high rate of false alarms while imposing complex and heavy structures for identifying TD~\cite{ali2012application,graf2010speeding,tsantalis2011identification,tsantalis2015assessing}.


\begin{wraptable}{r}{7.5cm}
\vspace{-15pt}
\scriptsize
\caption{Examples of SATD comments.}\label{tab:satd_examples}
\vspace{-5pt}
\setlength\tabcolsep{2pt}
\begin{tabular}{l|l}
     \textbf{Project}  &    \textbf{SATD comments}    \\\hline
Apache Ant &  // \texttt{cannot remove underscores due to} \\
& \texttt{protected visibility $>$:(} \\
EMF & // \texttt{TODO Binary incompatibility;} \\ 
& \texttt{an old override must override putAll.} \\
JFreeChart & // \texttt{do we need to update the crosshair values?} \\
JMeter &     // \texttt{Can be null (not sure why)} \\
SQuirrel & // \texttt{is this right???} \\  
ArgoUML & // \texttt{Why does the next part not work?}
\end{tabular}
\vspace{-10pt}
\end{wraptable} Therefore, several researchers proposed to target self-admitted technical debt identification as the first step since they are often intentionally documented or ``self-admitted'' (via source code comments) by the developers. Some examples of SATDs within the data are shown in Table~\ref{tab:satd_examples}. In summary, identifying and resolving SATDs have several benefits: 
\bi
\item Removing  SATDs early reduces the maintenance cost of a software project. As reported by \citet{wehaibi2016examining} and \citet{wang_attention_satd}, these SATDs have negative implications on the software development process, in particular by making it more difficult to change in the future.
\item With SATDs elimination, software projects have better evolvability trajectory for accelerating new functionalities addition and integration.
\item We can leverage those easily found SATDs as cheap training data for recognizing TDs~\cite{jitterbug}. SATDs are the documents of TDs that have been ``admitted'' by the
developers, so they are not a specific type of TDs. SATDs  cover different types of TDs such
as code, defect, and requirement debts by Bavota et al.'s categorization~\cite{bavota2016large}. In other words, as long as the document refers to some aspect of technical debt it is treated as SATD. According to the recent TDs categorization study, \citet{Fucci_MSR_21} showed how SATDs are mapped across 10 categories, e.g., poor implementation choices, partially implemented, functional issues, etc.
\ei


\begin{figure*}[!t]
\vspace{-5pt}
\centering \includegraphics[width=\linewidth]{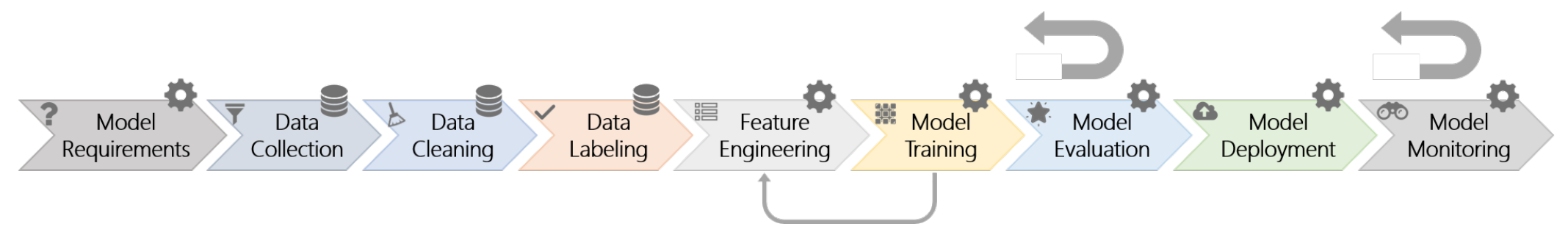}
\vspace{-15pt}
\caption{Nine stages of the machine learning workflow from a case study at Microsoft by Zimmermann et al. \cite{amershi_ICSE19AI4SE}. Some stages are data-oriented (e.g., data collection, cleaning, and labeling) and others are model-oriented
(e.g., model requirements, feature engineering, model training, evaluation, deployment and monitoring).}\label{fig:pipeline}
\vspace{-15pt}
\end{figure*}   

\subsection{Methods for Identification of Technical Debt}
One of the goals of industrial analytics is that new conclusions can be quickly obtained from new data just by applying data mining
algorithms. As shown in Figure \ref{fig:pipeline}, there are at least nine separate
stages that must be completed before that goal be reached~\cite{amershi_ICSE19AI4SE}. Each of these stages offers
unique and separate challenges, each of which deserves extensive attention. Many of these steps have been extensively studied in
the literature \cite{de2015contextualized,de2016investigating,de2020identifying,huang2018identifying,liu2018satd,maldonado2015detecting,maldonado2017using,potdar2014exploratory,jitterbug,zampetti2019automatically}. However, the labeling work of  step 4 has been receiving scant attention. In literature, there are several approaches for executing the labeling process:
\be
\item Manual labeling;
\item Crowd sourcing;
\item Reuse of labels;
\item Automatic labeling;
\item Active learning (which is a special kind of semi-supervised learning)
\ee

All of these approaches have their drawbacks; e.g. they are error-prone or will not scale. In response to these shortcomings, this study will take two directions:
\bi
\item First, we will try a {\em label-free} approach using  a combination of pure {\em unsupervised learning} techniques to pseudo-label the data, and subsequently {\em active} learning, i.e., \IT (0);

\item If the {\em label-free} approach fails, then we will try  a hybrid of an {\em active-learning} approach, called \IT{}(100), which starts with the help of {\em unsupervised learning} to first filter out the highly technical-debt prone comments before the  incrementally learning on all of the SATDs. 
\ei


\subsubsection{Manual labeling}

In  manual labeling, a team of (e.g.) graduate students assigns labels then (a) cross-checks their work via say, a Kappa statistic; then (b)~use some  skilled third person to resolve any labeling disagreements~\cite{lutz2004empirical,tu2020changing,tu2020better}.
 
Manual labeling   can be very slow. Tu et al. recently studied a corpus of 678 Github projects~\cite{tu2020better,tu2020changing}.
A  random selection of 10 projects from that corpus had
   $22,500$  commits, which took 175 hours to manually label the commits {\em buggy, non-buggy} (time includes cross-checking). That is, in a hypothetical situation of manual labeling
   500 projects (with each project has 5,000 commits) would  have
   required 90 weeks of   work.
   
   \subsubsection{Crowd Sourcing}
   Tu et al. \cite{tu2020better} offers a cost estimate of what resources would be required to sub-contract that effort to   dozens of  crowd sourced workers via tools like  Mechanical Turk (MT). Applying best practices in crowd sourcing~\cite{chen2019replication}, assuming (a)~at least USA minimum ages~\cite{silberman2018responsible};
   and (b)~our university taking a 50\% overhead tax on grants; then   crowd sourcing the labeling  of the issues from 
500 projects  would require   \$320,000 of grant reserve.

\subsubsection{Reusing Labels}

Because manual labeling can be time-consuming, crowd sourcing too expensive, and micro-labeling error-prone, researchers often reuse labels from previous studies ~\cite{xia16ist17,yang2015deep, yang16unsupervised}.
This approach is unsatisfactory for two reasons.
One, when exploring a new domain, there may be no
relevant, pre-existing labels to reuse. Two, reusing labels might propagate unsatisfactory label instances for future work. For example, the widely cited NASA datasets in defect prediction were found to have dubious quality~\cite{data_jinx, quality} in 2013 but has been utilized since then. Specifically,~\citet{jitterbug} were exploring self-admitted technical debt and found that their classifiers had an alarming high false positive rate. But when they manually checked the labels of their data taken from a prior study by~\citet{maldonado2015detecting}, they found that over 98\% of the reused false-positive labels were incorrect. Table~\ref{tab:updated_examples} shows some example
comments whose labels were updated in \citet{jitterbug}'s study.

\begin{table}[!tbh]
\scriptsize
\renewcommand{\arraystretch}{1.1}
\caption{ Examples of different labels from the original datasets curated by~\citet{maldonado2015detecting} and the updated datasets by~\citet{jitterbug}
} \vspace{-5pt}
\centering
\setlength\tabcolsep{4pt}
\begin{tabular}{p{.12\linewidth}|p{.52\linewidth}|P{.13\linewidth}|P{.13\linewidth}}
\textbf{Project} &  
\textbf{Comment Text} & 
\textbf{Original}  &
\textbf{Yu et al.'s} \\
 & & 
\textbf{Label~\cite{maldonado2015detecting}}  &
\textbf{Label~\cite{jitterbug}} \\
\hline
Apache Ant & 
\texttt{//TODO Test on other versions of weblogic //TODO add more attributes to the task, to take care of all jspc options //TODO Test on Unix} & 
no & 
yes \\
\hline
ArgoUML & 
\texttt{// skip backup files. This is actually a workaround for the cpp generator, which always creates backup files (it's a bug).}
 & no & yes \\
\hline
JFreeChart & 
\texttt{// FIXME: we've cloned the chart, but the dataset(s) aren't cloned and we should do that}  & 
no & 
yes \\
\hline
JRuby & 
\texttt{// All errors to sysread should be SystemCallErrors, but on a closed stream Ruby returns an IOError.  Java throws same exception for all errors so we resort to this hack...} & 
no & 
yes \\
\hline
Columba & 
\texttt{// FIXME r.setPos();} & 
no & 
yes \\

\end{tabular}
\vspace{-15pt}
\label{tab:updated_examples}
\end{table}

\subsubsection{Automatic labeling} 

If labels cannot be generated manually or reused from other papers, using automatic labeling processes is an attractive alternative. For example,  defect prediction papers~\cite{catolino17_jitmobile,hindle08_largecommits,kamei12_jit,Kim08changes,mockus00changeskeys,nayrolles18_clever,commitguru}  can  label a commit as ``bug-fixing''
 when the commit text contains certain  keywords (e.g.
"bug", ``fix", ``wrong", ``error", ``fail'', etc~\cite{tu2020better}).
 Vasilescu et al. \cite{Vasilescu18z,Vasilescu15github}  noted that these keywords are used in a somewhat ad hoc manner
(researchers peek at a few results, then tinker   with  regular expressions that combine these keywords). 
Tu et al. \cite{tu2020better} had found that these
simplistic keyword approaches can introduce many errors,
perhaps due to the specialization of the project nature or the ad-hoc nature of their creation ~\cite{Vasilescu18z}. 

Again, TDs are often ``self-admitted'' by developers in code comments~\cite{potdar2014exploratory} as shown in Table~\ref{tab:satd_examples}
in order to signal other developers that the corresponding code has \respto{3c} \revised{TD} that will be resolved for better results. In 2014, after studying four large-scale open-source software projects, Potdar and Shihab~\cite{potdar2014exploratory} concluded that developers may intentionally leave traces of TDs (i.e., SATDs) in their comments, such as ``{\em hack, fixme, is problematic, this isn't very solid, probably a bug, hope everything will work, fix this crap}''). 
These comments tend to  make SATDs much easier to find. Identifying and tracking SATDs have three important benefits as indicated \S2.1. There are two prominent approaches to automatically identify SATDs:

{\bf 1. Pattern-based approaches}~\cite{de2015contextualized,de2016investigating,de2020identifying,maldonado2015detecting,potdar2014exploratory} consist of three steps: (1) manually inspect code comments and label each one as SATD or non-SATD, (2) manually analyze the labeled items and summarize patterns for SATDs, e.g., if a comment has keywords like ``{\em hack, fixme, probably a bug}'', then it has a high chance of being related to a SATD, (3) apply the summarized patterns to unlabeled comments to identify SATDs. Instead, \citet{jitterbug} proposed  Easy as the  SOTA pattern-based method to automatically identify 20-90\% of SATDs by finding patterns associated with high precision (close to 100\%).

\noindent \underline{\textit{Limitation of the SOTA Pattern-based approach}}: this  approach does need
extensively labeled
training data   to find   patterns that are associated with SATDs because it relies on precision. As mentioned above, generating that data  requires intensive labor and expensive cost. Moreover, this method can still miss up to 80\% of SATDs.

{\bf 2. Machine Learning approaches}~\cite{huang2018identifying,liu2018satd,maldonado2017using,zampetti2019automatically} involve models working in the \textit{supervised learning} manner, which are trained on labeled SATD datasets to learn the underlying rules of comments admitting TDs.
For example, Tan et al.~\cite{tan2007icomment,tan2012tcomment} analyzed source code comments using natural language processing to understand programming rules and documentations and indicates comment quality and inconsistency. 
In 2017, Maldonado et al.~\cite{maldonado2017using} successfully identified two types of SATD in 10 open-source projects (average 63\% F1 Score) using Natural Language Processing (Max Entropy Stanford Classifier) using only 23\% training data. 
Huang et al.~\cite{liu2018satd} introduced a Multinomial Naive Bayes sub-classifier for each training dataset using information gain as feature selection then combine those sub-classifiers with boosting technique to achieve an average of 73\% F1 scores ~\cite{huang2018identifying}. 
A recent IDE for Eclipse was also released using this technique for identifying SATD in Java projects~\cite{liu2018satd}. 
Recently, some studies explore different feature engineering for identifying SATDs, e.g. Wattanakriengkrai et al.~\cite{wattanakriengkrai2019automatic} applied N-gram IDF as features, and Flisar and Podgorelec~\cite{flisar2019identification} explored how feature selection with word embedding can help the prediction. The latest progress are from Wang et al.~\cite{wang_attention_satd}'s HATD and Ren et al.~\cite{ren2019neural}'s tuned CNN utilized a deep convolutional neural network to achieve a higher F1 score than all the previous solutions. The HATD paper asserts that their algorithm
defeats CNN but, after much effort, we could not reproduce that result\footnote{We found that there is no  reproduction package published with HATD. We tried contacting the authors of that paper, without success.}.  These machine learning models can be a good indicator for which comments are more likely to be related to SATDs. 

\noindent \underline{\textit{Limitation of the SOTA Machine Learning approach}}: deep learners often require having access to a substantial amount of labeled data which is not always available, especially in new domains (e.g., the success of open-source projects). With precision ranging from 60\% to 85\%, it is not reliable to fully automate the process. Human experts are then required to verify every decision the machine learning model made and thus costs a large amount of time and labor.


\subsubsection{Semi-supervised Learning}\label{actlearn}

Finally, another approach is to only label a representative sample of the data, build a classifier from that sample, then use that classifier to label the remaining data~\cite{ref49}. To find that representative example, an unsupervised learner (e.g. associations rule learner), a clustering algorithm, or an instance selection algorithm is used to find repeated patterns in the data~\cite{Kim08changes}. Then a human oracle is asked to label one exemplar from each pattern. More sophisticated versions of this scheme include \emph{active learners}, where an AI tool advances ahead of the human to fetch the most informative next sample to be labeled~\cite{kocaguneli2012active,settles2009active}.
If humans agree to first label only the most informative examples, then active learners can be used to produce better models more efficiently by reducing the number of examples that humans have to label.

The more general term for {\em active learning} is {\em semi-supervised learning}. Both terms mean ``do what you can with a small sample of the labels'' while {\em active learning} adds a feedback loop that checks new labels one at a time with an oracle. Moreover, semi-supervised learning relies on partially labeled data and mostly unlabeled data.

Since 2012, active learning approaches have received scarce attention in SE~\cite{kocaguneli2012active,tu2020better,yu2019improving,jitterbug}.
Initially, active learning seems to be a promising method for addressing the cost of label checking and generating:
for self-admitted technical debt, only 24\% on a median of the training corpus had to be labeled~\cite{jitterbug}; using active learning, effort estimation for $N$ projects only needed labels on 11\% of those projects~\cite{kocaguneli2012active}; further, 
while seeking 95\% of the vulnerabilities in 28,750 Mozilla Firefox C and C++ source code files, humans only had to inspect 30\% of the code~\cite{yu2019improving}. However, active learning still produces disappointing results. For example, it is still a daunting task to ``only'' label 5\% to 10\% of the projects in the 1,857,423 projects in RepoReapers~\cite{curating} or the 
9.6 million links explored by~\citet{Hideaki19}. Although it might be justified for very mission-critical projects, consider the Firefox study~\cite{yu2019improving} which required the human effort of inspecting 28,750 (total source code files) x 30\% = 8,625 source code files to identify 95\% of the vulnerabilities. This is beyond the resources of most analysts.

Several two-step frameworks were proposed for the active learning approach. \citet{jitterbug} proposed Jitterbug to identify SATDs: (1) identify patterns for the ``easy to find'' SATDs (20-90\% of all SATDs) with close to 100\% precision and automatically classify comments with the patterns as SATDs (without human verification), (2) apply machine learning techniques to guide human experts to find the remaining ``hard to find'' SATDs with least number of comments read. Interestingly,~\citet{guo2019mat} utilized a similar idea but using only four keywords ({\em ``fixme, todo, hack, xxx''}) to identify the ``easy to find'' SATDs and applied supervised learning models to incrementally find the remaining ``hard to find'' SATDs.

\noindent \underline{\textit{Limitation of the SOTA Active Learning approach}}: 
\bi
\item Costly in real-world: both steps (pattern-based method and machine learning) require the training data to be labeled in order to proceed which can be costly in the real world, especially in a new domain. More importantly, Jitterbug's active learning strategy relies on the first step of the pattern-based method in order to reach the target recall. Thus, without the labeled training data, Jitterbug's guarantee in reaching the user-specified recall would be almost impossible.
\item Difficult for Active Learning: the first step of the pattern-based approach identified up to 90\% of the SATDs, but this makes it difficult for the active learning strategy to find the rest of 10\% SATDs. This can increase human effort to review the labels. This will be confirmed later in \S\ref{tion:threats}. 
\ei 

\begin{table}[!t]
\vspace{-10pt}
\setlength\tabcolsep{2.5pt}
\renewcommand{\arraystretch}{1.5}
\caption{Differences between two SOTAs of \citet{ren2019neural} and \citet{jitterbug} against our work, DebtFree. Again, the first two steps of DebtFree are optional as they will be investigated later.}
\vspace{-15pt}
\label{tbl:diffs}
\scriptsize
\begin{center}
\begin{tabular}{p{1.2cm}|p{3.1cm}|p{3.1cm}|p{3.1cm}}
& \textbf{\citet{ren2019neural}}   & \textbf{~\citet{jitterbug}} & \IT{} \\ \hline
\textbf{Learning Type}  & Supervised  & Semi-Supervised  & Semi-Supervised   \\ \hline
& 
1. Trains the deep learning model CNN on $(N-1)$ labeled datasets.  & 
1. Filtering out easy SATDs in the target $i$ dataset with the pattern recognizer, Easy, to learn patterns with higher than 80\% precision on $(N-1)$ labeled datasets.   &
1. (Optional) Pseudo-labels the $(N-1)$ datasets with the unsupervised learner, CLA~\cite{nam2015clami}. \\ 
\textbf{Core Process} & 2. Uses the trained CNN model to identify SATDs on the target $i$ dataset  & 
2. Trains the RF model on $(N-1)$ labeled datasets and incrementally update the model in the active learning manner.   &
2. (Optional) Filtering out easy SATDs in the target $i$ dataset with the pattern recognizer (e.g., Easy or CLA) on $(N-1)$ pseudo-labeled/labeled datasets. \\ 
&  &   &
3. Trains the RF model on $(N-1)$ pseudo-labeled/labeled datasets and incrementally update the model in the active learning manner. \\ 
\hline
& - Use all 100\% labels from $(N-1)$ datasets. & - Use all 100\% labels from $(N-1)$ datasets. & - Use 0\% labels from $(N-1)$ datasets with step 1. \\
\textbf{Labeling Effort}   & - No labeling done on the target $i$ dataset. 
& - On average, labeling 23\% on the target $i$ dataset. 
& - On average, labeling 11\% on the target $i$ dataset. \\

\end{tabular}
\end{center}
\label{tab:differences} 
\vspace{-25pt}
\end{table}

Hence, in this SATDs identification work, we aim to reduce the labeling cost of both SOTA works including the SOTA semi-supervised learner from \citet{jitterbug} and the SOTA supervised learner from \citet{ren2019neural}. In the process of developing such a method, our investigation shows our proposed method, \IT{}, also performs statistically similar or even better. The differences between our approach and their are described in Table~\ref{tab:differences}. The investigation explores the usefulness of \textit{unsupervised learning} through a mix of approaches: (1) unsupervised learning in low-resource setting (unlabeled data) can frugally pseudo-label the training data, (2) an unsupervised learner acts as a preprocessor to filter out SATDs without relying on data labels in high-resource settings (labeled data), and (3) a tuned active learning strategy to specialize the learning on the target dataset by filtering out the training datasets when there is no benefit from learning on them anymore.

In order to overcome the previously documented limitations in identifying SATDs (i.e., the previous supervised learning and active learning methods are expensive), unsupervised learning is a promising direction, but not competent enough. To address this literature gap, it is an opportune time to propose the \IT{} framework, which is based on the integration of unsupervised learning and active learning. \IT{} investigates the combination of three approaches including pseudo-labeling, filtering, and active learning with different candidates for each approach. \IT{}'s configurations are formulated after exploring two scenarios: training data labels are known and training data labels are unknown. In the case of low-resource (training data with no labels), we will pick \IT{}(0). On the other hand, given resources (training data with labels), \IT{}(100) is employed instead.

\section{Methodology} \label{sec:methodology}

\subsection{General Framework}

Our proposed \IT{} is an end-to-end solution that labels the data, extends the data corpus, and identifies SATDs in a semi-supervised learning approach. \IT{} is comprised of two settings \IT{}(0) and \IT{}(100).


\subsection{DebtFree(0)}

When there is no access to the labels of the training data, our study shows that the filtering step is not needed here and the best combination for \IT{}(0) consists of two steps: 
unsupervised learning with CLA~\cite{nam2015clami} to cheaply pseudo-labels the training data, and then our proposed active learning strategy, Falcon, to incrementally update and learn to identify the SATDs on the test data. 




\subsubsection{Pseudo-labeling via CLA/CLAFI} \label{sec:clami}

In the SOTA literature and comparative study of unsupervised models in defect prediction, 
CLA starts with two steps of (1) \underline{\textbf{C}}lustering the instances and (2) \underline{\textbf{LA}}beling those instances accordingly to the cluster. In the low resource setting with no labels available, we can label/predict all instances. CLAFI is an extension of CLA which is a full-stack framework that also include (3) \underline{\textbf{F}}eatures selection and (4) \underline{\textbf{I}}nstances selection. Both CLA and CLAFI were first proposed by~\citet{nam2015clami} in the domain of defect prediction. The intuition of such methods is based on the defect proneness tendency that is often found in defect prediction research, that is \textit{the higher complexity is associated with the proneness of the defects} \cite{DAmbros2012,hassan_icse09, menzies07defect, nam2015clami, Rahman:2013,  tu2021frugal}. Put simply, there is a tendency where the problematic instance's feature values are higher than the non-problematic ones. For instance, Hassan et al.~\cite{hassan_icse09} predicted defects using the entropy (or complexity) of code changes (the more complex changes to a file, the higher the chance the file will contain faults). This tendency and CLA's/CLAFI's effectiveness were confirmed via the recent literature and comparative study of 40 unsupervised models in defect prediction across 27 datasets and three types of features. They found CLA's/CLAFI's performance is superior to other unsupervised methods while similar to supervised learning approaches. Moreover, Tu et al.~\cite{tu2021frugal} recently applied this intuition in developing their method to further the SOTA work for static analysis and issue close time prediction. Therefore, this study investigates and finds that the hypothesized tendency is also applicable in SATDs data but only effective for the semi-automated method but not the fully automated one. CLA is preferred over CLAFI but this study examines both before choosing one over the other.

Before CLA or CLAFI, \IT{}(0) extracts features from each comment candidate as $L2$-normalized terms (the square root of the sum of the squared vector) terms with the  
TF-IDF\footnote{For token $t$, its tf-idf score: $\mathit{Tfidf}(t) = \sum_{d\in D} \mathit{Tfidf}(t,d)$, in which for token $t$ in comment or document $d$, $\mathit{Tfidf}(t, d)=w^t_d\times (\log \frac{|D|}{\sum_{d\in D} \mathit{sgn}(w^t_d)}+1)$ where $w^t_i$ is the number of times token $t$ appears in document $d$.} scores (after stop word removal).

\noindent \underline{\textbf{Clustering}}: 
\be
\item Find the median of feature $F_1, F_2,..., F_n$ ($median(F_i)$) across the whole dataset.
\item For each data instance $X_i$, go through each feature value of the respective data instance to count the time when the feature $F_i > median(F_i)$ as $K_i$.
\ee

\noindent \underline{\textbf{Labeling}}: label the instance $X_i$ as SATD if $K_i > median(K)$, else label it as non-TD.

\noindent \textbf{\underline{Feature Selection}}: Calculate the violation score per feature, called metric in the original proposal of Nam et al. \cite{nam2015clami}. The process is done on both the train and the test dataset. 
\be
\item For each $F_i$, go through all instances of $X_j$, a violation happens when $F_i$ at $X_j$ is higher than the $median(K_i)$ but $Y_j=1$ and vice-versa. 
\item Sum all the violations per feature across the whole dataset and sort it in ascending order.
\item Select the feature with the lowest violation score, if multiple of them have the same score then pick all of them. 
\ee 

\noindent \textbf{\underline{Instance Selection}}: 
\be
\item With the selected features, go through each instance $X_i$ and check if the respective $F_j$ values violated the proneness assumption then remove that instance $X_i$. 
\item If the dataset does not have instances with both classes at the end then pick the next minimum violation score to select metrics. 
\item This process is only done on the training dataset.
\ee

After selecting features with the minimum violation scores and removing the instances that violated the technical-debt proneness tendency, a practitioner can train any machine learners on the preprocessed training data to identify the SATDs from the unlabeled/test dataset. For this step, we picked Random Forest which is also Jitterbug's choice of learner after being compared across other ones~\cite{jitterbug}.

\subsubsection{Active Learning via Falcon} \label{sec:cl_0}

The data that do not share the technical-debt proneness tendency is being continuously learned through human and AI partnership, named Falcon as the proposed active learning strategy from this study. First, a classification model (for SATD
or non-SATD comments) is trained on the training dataset. When reading the new comments, Falcon
initially uses uncertainty sampling to quickly build the model, then switches to certainty sampling to greedily find technical debt comments. The machine learner (RF) uses this feedback from human to learn their models incrementally. These trained model then sorts the stream of comments such that humans read the most informative ones first
(and the comments are sorted again each time a human offers a new label for a comment). After the found SATDs reach a specific threshold then Falcon drops the training data and incrementally updates it with the target data. This is similar to the separation action of the Falcon rocket to drop the thruster for boosting up the rocket's speed after reaching a required height. More specifically, Falcon executes as follows: \\
\begin{enumerate}[start=1,label={\bfseries Step \arabic*}]
\item \textbf{Feature Extraction:} Given a set of comments candidates, Falcon extracts features from each candidate as $L2$-normalized  terms with the highest 
TF-IDF. Initialize the set of labeled data points as $L \leftarrow \emptyset$ and the set of labeled positive data points as $L_B \leftarrow \emptyset$.
\item \textbf{Bootstrap Learning:} Falcon utilizes the labeled training dataset to train a machine learning model, i.e., Random Forest (Yu et al.'s choice of learner \cite{jitterbug}). 

\item \textbf{Initial Sampling:}
Falcon starts by
randomly sampling unlabeled candidate studies until humans declare that they see $N_1=1$ technical-debt examples. 
\item \textbf{Uncertainty Sampling:}
Then, as human assessors offer labels, one example at a time, Falcon
trains and updates with weighting to control query with uncertainty sampling, until $N_2=10$ technical-debt examples are found.
Here, different weights are assigned to each class
($W_B= 1/|L_B|$, $W_N= 1/|L_N|$).
\item \textbf{Certainty Sampling:}
Next, Falcon trains further using
certainty sampling and 
Wallace's ``aggressive undersampling'' \cite{wallace2010semi}
that culls majority class examples closest to the decision boundary.
\item \textbf{Training Data Separation:} Falcon drops training data after finding more than 10\% of estimated SATDs then the model is retrained on the reviewed target data.
\item \textbf{Early Stopping:} Falcon
stops training when it is estimated that \mbox{$N_3=90\%$} of the SATDs have
been found.
\end{enumerate} 
To generate the $N_4$ estimate,
whenever the RF model is retrained, Falcon makes temporary ``guesses'' about the unlabeled examples (by running those examples through the classifier).
To turn these guesses into an estimate of the remaining technical-debt comments, Falcon:\\
\begin{enumerate}
\item Builds a fast and simple model, e.g., Logistic Regression, using the guesses. Faster feedback cycle to update the model and for the users to make decisions
\item
Using that regression model, Hard makes new guesses on the remaining unlabeled examples.
\item
Loops back to step1 until the new guesses are the same as the guesses in the previous loop.
\item
Uses this logistic regression model to estimate the remaining number of positive examples in the data. 
\end{enumerate}

The reader will note that there are many specific engineering decisions built into the above
design (e.g. the values $\{N_1=1, N_2=10, N_3=90\%\}$). Those decisions were initially made by 
Yu et al.~\cite{Yu2018} after exploring 32 different kinds of active learners. Those were later adopted by the SOTA active learning framework Jitterbug~\cite{jitterbug} for SATD identification. 

\subsection{DebtFree~(100)}


\IT{}(100) is inspired by the SOTA's Jitterbug framework in the high-resource setting, i.e., have access to labeled training data. Instead of a pattern-based approach with a active learning strategy, we propose to replace the pattern-based approach with an unsupervised learner to filter out the highly technical-debt prone comments. Then, the state-of-the-art active learning approach (i.e., Hard~\cite{jitterbug}) guides the human experts to first start with training on training data labeled by the experts and then identify all of SATD comments on new coming data. 

\subsubsection{Filtering via CLA}

Here, SATDs from the test data can be early filtered out with CLA. However, instead of picking $threshold_K = median(K)$ to label the training and test data, we iterate $threshold_K = percentile(all_K, i)$ for i from 50\% to 95\% percentile. Then, the best $threshold_K$ is the one with the highest precision to identify the SATDs within the training set. The test set with $K_i > best\_threshold_K$ are labeled as SATDs. Then, instances that meet the tendency in both test and training datasets are removed. The intuition is similar to Yu et al. \cite{jitterbug} is that the ``easier'' SATDs can first be easily identified where easier here means they share the tendency of \textit{the higher complexity is associated with the proneness of the technical-debts}.

\begin{table*}[!b]
\large
\caption{Differences between the active learning strategies for this study: EMBLEM~\cite{tu2020better}, HARD~\cite{jitterbug}, and our proposed approach, i.e., FALCON.}
\begin{center}
\resizebox{0.7\linewidth}{!}{\renewcommand{\arraystretch}{2}%
\begin{tabularx}{\textwidth}{p{2in}|c|c|c}
& \textbf{ Emblem\cite{tu2020better}} & \textbf{Hard\cite{jitterbug}} & \textbf{Falcon} \\ \cline{1-4}
\multicolumn{1}{m{2in}|}{start with learning on existing training datasets.} & {\Large \textbf{\redx}} & {\Large \greencheck} & {\Large \greencheck} \\ \cdashline{1-4}
\multicolumn{1}{m{2in}|}{drop the training data and retrain the model on the reviewed test data after attaining a certain threshold of SATDs } &  {\Large \redx} & {\Large \redx} & {\Large \greencheck} \\ \cline{1-4}
\end{tabularx}}
\end{center}
\label{tab:difference}
\end{table*}

\subsubsection{Active Learning via Hard} \label{sec:continuous_learning}

\begin{figure*}[!t]
\minipage{0.34\textwidth}%
\vspace{45pt}
\begin{center}
  \scalebox{.75}{\includegraphics[width=\linewidth]{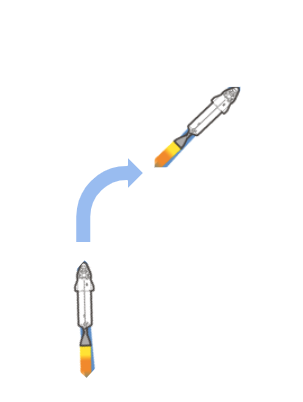}}
\end{center} 
\vspace{-11pt}
\subcaption{Emblem}
\endminipage
\minipage{0.34\textwidth}
\begin{center}
\scalebox{.85}{\includegraphics[width=\linewidth]{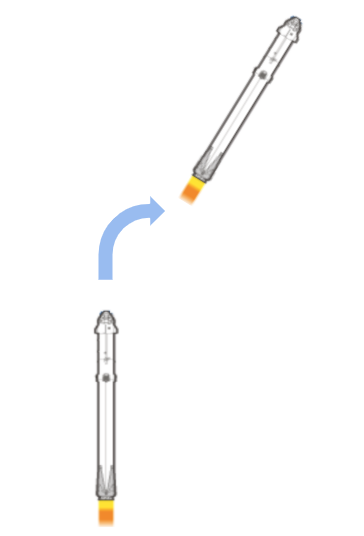}}
\end{center}
 \subcaption{Hard}
\endminipage
\minipage{0.34\textwidth}%
\vspace{-10pt}
\begin{center}
  \scalebox{.75}{\includegraphics[width=\linewidth]{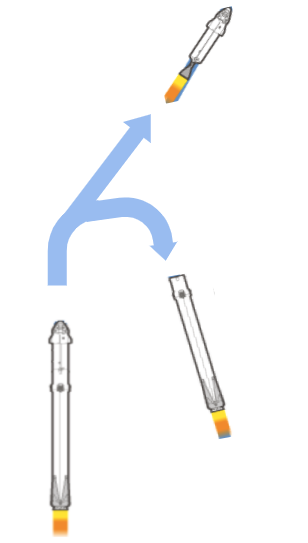}}
\end{center}
\subcaption{Falcon}
\endminipage
\caption{The visual analogy of rocket launching for the comparison of Emblem, Hard, and Falcon processes. 
}
\label{fig:analogy}
\end{figure*}
The active learning strategy here is Hard from~\citet{jitterbug}'s Jitterbug which is similar to Falcon (as discussed previously in \S\ref{sec:cl_0}) except it does not filter out training data at any point in the learning process. Another variation for comparison also includes Tu et al.'s Emblem~\cite{tu2020better}. Emblem only uses one hundred randomly sampled labeled test data to start the active learning strategy. For Emblem, we suspected that due to the randomness in sampling, in the first one hundred instances the performance will be unstable. In summary, the differences and similarities are documented in Table~\ref{tab:difference} and the performance of all three methods will be compared in RQ2.1. For a more intuitive comparison, Figure~\ref{fig:analogy} shows the rocket launching process in each method. Emblem is cheap with no need for the booster (labeled training data) but difficult to reach a substantial height (acquiring SATDs). Hard is effective but expensive to carry the whole booster all the way till the end. Falcon is a hybrid that uses the booster upon launch to boost the rocket's speed but is discarded once the rocket reaches a certain height.




\section{Experimental Design} \label{sec:experiment}
\subsection{Data}
\label{sec:case_study}
 
To validate this study's hypothesis and the proposed methods' effectiveness, we use the dataset from Maldonado et al. \cite{maldonado2015detecting} which has been corrected by Yu et al. \cite{jitterbug}. The dataset includes ten open-source projects collected from Github: Apache-Ant-1.7.0, Apache-Jmeter-2.10,
Hibernate-Distribution3.3.2.GA, ArgoUML, Columba-1.4-src, EMF-2.4.1, jEdit-4.2, jFreeChart-1.0.19, jRuby-1.4.0, SQL12. The provided dataset contains project
names, classification type (if any) with actual comments.

Table \ref{tab:details} lists the varying statistics of the comments found in these 10 projects. As only a small ratio of the source code comments describe SATDs, it would be time-consuming to label all comments manually. Thus, Maldonado and
Shihab developed 5 filtering heuristics to eliminate comments that are unlikely to be classified as
comments \cite{maldonado2015detecting}: (1) remove license comments/auto-generated comments, (2) remove commented source code, (3) remove Javadoc commented source comments, (4) group multiple single-line comments together, and (5) removing duplicate comments. In the end, the number
of comments that require manual annotation was significantly reduced from 259,229 to 62,275 (reducing the data by 75\%).

\begin{table}[!t]
\footnotesize
\caption{Dataset Details}
\centering
\begin{tabular}{p{.11\linewidth}|p{.09\linewidth}|p{.15\linewidth}|P{.13\linewidth}|P{.135\linewidth}|P{.135\linewidth}}
\textbf{Project} & 
\textbf{Release / Year} & 
\textbf{Domain} & 
\textbf{Comments} &
\textbf{Original SATDs~\cite{maldonado2015detecting}} &
\textbf{Corrected SATDs~\cite{jitterbug}}\\
\hline
Apache Ant & 
1.7.0 / 2006& 
Automating Build&
4098 &
131 (3.2\%) & 135 (3.3\%)\\
\hline
JMeter & 
2.10 / 2013 & 
Testing &
8057 &
374 (4.6\%) & 416 (5.2\%)\\
\hline
ArgoUML & 
- & 
UML Diagram &
9452 &
1413~(15\%) &  1630~(17.3\%)\\
\hline
Columba & 
1.4 / 2007 & 
Email Client &
6468 &
204 (3.2\%) & 220 (3.4\%)\\
\hline
EMF & 
2.4.1 / 2008& 
Model Framework &
4390 &
104 (2.4\%) &  119 (2.7\%)\\
\hline
Hibernate & 
3.3.2 / 2009 & 
Object Mapping Tool &
2968 &
472 (16\%) & 493 (17\%)\\
\hline
JEdit & 
4.2 / 2004 & 
Java Text Editor &
10322 &
256 (2.5\%) &  259 (2.5\%)\\
\hline
JFreeChart & 
1.0.19 / 2014 & 
Java Framework &
4408 &
209 (4.7\%) & 247 (5.6\%)\\
\hline
JRuby & 
1.4.0 / 2009 & 
Ruby for Java &
4897 &
622~(12.7\%) & 665~(13.4\%)\\
\hline
SQuirrel & 
- & 
Database &
7215 &
286 (4\%) &  313 (4.3\%)\\
\hline
\textbf{SUM} & 
 & 
 &
\textbf{62275} &
\textbf{4071~(6.5\%)} &
\textbf{4497~(7.2\%)} \\
\textbf{MEDIAN} & 
 & 
 &
\textbf{5683} &
\textbf{271~(4.8\%)} & \textbf{286~(5\%)}\\
\end{tabular}
\label{tab:details}
\end{table}

Furthermore in their work~\cite{maldonado2015detecting}, two humans then manually classified each comment according to the six different types of SATD mentioned by Alves et al.~\cite{alves2014towards} if they contained any SATD at all, else marked them $WITHOUT\_$ $CLASSIFICATION$.  Note
that, we do not use the
fine-grained SATD categories proposed by  \citet{maldonado2017using}, rather
we focus on a binary problem of instances being a SATD or not. Our study is concerned specifically with identifying if a problem is SATD 
 (e.g., $DEFECT$, $IMPLEMENTATION$, $DESIGN$, etc) or not ($WITHOUT\_$ $CLASSIFICATION$).  Simply, as long as the code comment refers to some aspect of technical debt it is treated as SATD. Similarly to previous work~\cite{jitterbug, ren2019neural},
we have changed the final label into a binary problem by defining
\textit{WITHOUT\_CLASSIFICATION} as $no$ and the rest of the categories as $yes$.  Stratified sampling of the dataset is applied to check personal bias, revealing with a $99\%$ confidence interval of $5\%$. 
A third human verified the agreement between the two using stratified sampling and reported a high level of agreement (Cohen's Kapp~\cite{cohen1968weighted} coefficient of $+0.81$). The significantly high level of agreement indicates that the dataset is unbiased and reliable. 

On the contrary, Yu et al. \cite{jitterbug} inspected this dataset for biases, and found that 98\% of the false positives were wrongly labeled. The differences between the original and corrected SATDs count are reported in the last two columns of the Table \ref{tab:details}. Specifically,  Maldonado et al. \cite{maldonado2015detecting} missed more than 10\% of the total SATDs. This discrepancy highlights the importance of validating prior research's conclusions, data, and methodologies should one employ them in their work, a process that might add significant overhead to the amount of required effort. Fortunately,  active  learning offers feasible remedial venues. Our proposed method DebtFree also established new state of the art that outshined the state-of-the-art active learning method for technical debt, Jitterbug.


\subsection{Data Miners} \label{sec:learners}

There are several data miner options for the active learning part of the \IT{}(100) or supervised learning part of the \IT{}(0). For this study, we only test simple and fast learners since the active learning model is updated/re-trained frequently while practitioners appreciate quick feedback loop to improve software,code,  and technical-debts. Such learners include:

\textbf{Logistic Regression:} a statistical method that uses a logistic function to model a binary dependent variable~\cite{hosmer2013applied}. A standard logistic function is a common ``S'' shape with Equation~\ref{eq:LR}:
\begin{equation}
\label{eq:LR}
p(x) = \frac{1}{1+e^{-(\beta_0+\beta_1x)}}
\end{equation}
where $p(x)\in [0,1]$ for all $t$. Fitting on the training data, logistic regression looks for the best parameter $\beta$ to classify input data $x$ into two target classes $\{0,1\}$. LR is used for estimating SATDs in \IT{}(100)'s active learning.

\textbf{Random Forest:} an ensemble learning method that constructs a multitude of decision trees,
each time with different subsets of the data rows $R$ and columns $C$\footnote{Specifically, using $\log_2{C}$ of the columns, selected at random.}.
Each decision tree is recursively built to
find the features that yields the most reduction in {\em entropy} (higher entropy indicates lower ability to draw conclusions in the paritioned data)~\cite{CART}.
Test data is then passed across all $N$ trees. Conclusions are determined by a majority vote across all the trees~\cite{Breiman2001}. Holistically, RF is based on bagging (bootstrap aggregation) which averages the results over multiple trees from sub-samples (reducing variance). Both methods are popular in Machine Learning, and are implemented in the open-source toolkit Scikit-learn \cite{pedregosa2011scikit}. Random Forest is used for classifying SATD comments in \IT{}(0), \IT{}(100), and Yu et al.'s Jitterbug \cite{jitterbug}.

\subsection{Experimental Process}

Experiments are conducted on the SATD dataset with 10 projects described in \S\ref{sec:case_study}. Each time, one project is selected as a target project (with labels unknown) and the rest 9 datasets are treated as source projects. This study addresses the following research questions:

\textbf{RQ1: How well can the state-of-the-art unsupervised learning method identify SATDs?} 

\textbf{RQ2: How can the state-of-the-art active learning framework be combined with the state-of-the-art unsupervised learner?}

\textbf{RQ3: How does the proposed \IT{} perform against state-of-the-art models in identifying SATDs?}

Recall our study investigates the combination of three approaches including Pseudo-labeling, Filtering, and Active Learning. In this big data era, there are available datasets easily curated from the web (e.g., Github) with no labels and we can harvest valuable insights from them without labeling the data. The first RQ demonstrates that unsupervised learning by itself is not effective but it is useful for understanding some hidden patterns and early filtering out SATDs. In order to extend both the unsupervised learning approach and the active learning approach, the second RQ experiments different ways of integrating both frameworks in two settings with \textit{RQ2.1} where there is no access to the labels of the training data and \textit{RQ2.2} where there is access to the labels. For \textit{RQ2.1}, we test different candidates for all three steps whereas in \textit{RQ2.2} we only focus on candidates for the filtering step and the active learning step.  With RQ2.1, we can simulate a real world situation of having data with no labels by hiding the labels of the training data. During the active learning step across the experiments, the oracles are queried for the target project, the ground truth labels are applied to label the queried comments to simulate the human-in-the-loop labeling process.

From RQ2, we finalize the best combination for two settings from RQ2 as \IT{}(0) and \IT{}(100). In RQ3, we assess \IT{}(0)/(100)'s usefulness by comparing them  against the SOTA semi-supervised learning work  by \citet{jitterbug} (i.e., Jitterbug) and the SOTA supervised learning work  by \citet{ren2019neural} (i.e., CNN).  We recycle the available implementations of their approaches instead of re-implementing them ourselves to generate the results. Therefore, we are more confident that our conclusions or insights would align with the original work. \vspace{-5pt}

\subsection{Statistical Testing}

We employ Cohen'$d$ effect size test to determine which results are similar by calculating $medium\_{step2}$. We take guidance from \citet{Sawilowsky2009NewES} at al.'s work to determine the value of $d$ to be used.
That paper asserts that ``small'' and ``medium'' effects can be measured using $d = 0.2$ and $d = 0.5$, respectively. We analyze this data looking for differences larger than $d = (0.5 + 0.2)/2 = 0.35$. This $d$ is higher than the Jitterbug's $d = 0.2$, ensuring the differences in the results are statistically significant, and consequently higher confidence in the effectiveness of the proposed methods:
\vspace{-3pt}
\begin{equation}
\small
\label{eq:cohen_step2}
Medium_{step2} = 0.35 \cdot StdDev(\text{All results})
\end{equation}

\section{Results} \label{tion:result}

This section will provide details on the experiments and results for answering the research questions listed above.

\textbf{RQ1: How well can the state-of-the-art Unsupervised Learning method identify SATDs?}


In this experiment, we compare the performance of the following five treatments:
\bi
\item
\textbf{CLA:} The unsupervised learner described in \S\ref{sec:clami}, which clusters the test data instances based on the features' median by marking the ones with features higher than the median as SATDs and vice-versa. 
\item
\textbf{CLAFI+RF:} A combination of CLAFI (\S \ref{sec:clami}) and the data miner RF (\S\ref{sec:learners}). It started with pseudo-labeling the unlabeled training data with CLA. Then the data is preprocessed by removing the features (for nine train and one target datasets) and instances (only on the nine training datasets) that violated the assumption from CLA. Then we pick Random Forest as the learner choice from SOTA Yu et al.'s Jitterbug work's data miner of choice for supervised learning. 
\item
\textbf{CLA+RF:} This is similar to the previous step except no features and instances preprocessing step. 
\item
\textbf{Easy:} The pattern-based approach, first step of Jitterbug. For each pattern $p$, find $score = Prec(p)^4 \cdot P(p) = TP(p)^4 / P(p)^{3}$ where $P(p)$ is the number of comments $p$ (positives) and $TP(p)$ is the number of SATD comments containing $p$ (true positives). We iteratively select the pattern with the highest
score, then removes comments
containing that pattern from both train and test data until the selected pattern has lower than 80\% 
precision on the training data. Then, stop and evaluate. 

\item
\textbf{Easy+CLA:} First, the pattern-based approach Easy filters out test data comments with patterns that have more than 80\% precision on training data and then passed to CLA for identified the rest of SATDs. 

\ei

These pattern-based approaches can be seen as two groups: unsupervised learners (i.e.,  CLA variants) and supervised learners (i.e., Easy and Easy+CLA). In term of the effectiveness of these methods, it is recommended to evaluate multiple metrics so we employ Recall and APFD.





Recall measures the ability to identify the SATDs, $Recall = TP / (TP+FN)$. $TP$ is the number of true positives (SATD comments predicted as SATDs), $TN$ is the number of true negatives (non-SATD comments predicted as non-SATDs), $FP$ is the number of false positives (non-SATD comments predicted as SATDs), and $FN$ is the number of false negatives (SATD comments predicted as non-SATDs).

In order to take effort into consideration for the performance of the model, we also employ APFD. APFD calculates the area under the curve of the recall-cost curve whereas other metrics (e.g. precision, recall, F1, G1) only evaluate a single point of the curve. APFD ranges from 0.0 to 1.0, with higher values indicating that higher recall could be achieved at a lower cost, and thus, more advantageous.  An APFD value  of 0.5 can be achieved by randomly select the next item each time.

\bi 
\vspace{5pt}

\item \small $Recall = \frac{|\{\text{SATDs}\}\cap\{\text{human verified comments}\}|}{|\{\text{SATDs}\}|}$ \\
\item  \small $Cost = \frac{|\{\text{human verified comments}\}|}{|\{\text{comments}\}|}$ \\
\ei

\begin{table*}[!t]
\scriptsize
\caption{Comparison between \textbf{CLA}, \textbf{CLAFI+RF}, \textbf{CLA+RF}, \textbf{Easy}, and \textbf{Easy+CLA} are made in terms of Recall and APFD (the higher the better). Medians and IQRs (delta between 75th percentile and 25th percentile, lower the better) are calculated for easy comparisons. \textbf{CLA}, \textbf{CLA+RF}, and  \textbf{CLAFI+RF} do not have access to training data's labels while \textbf{Easy} and \textbf{Easy+CLA} do. Here,  the \colorbox[HTML]{D28986}{darker} cells show top rank while the \colorbox[HTML]{FFCCC9}{lighter} cells show rank two where the difference between two ranks is statistically significant by being higher than $M$ reported in the left most column.}
\centering
\setlength\tabcolsep{5pt}
\begin{tabular}{c|l|c|c|c|c|c|c|c|c|c|c|c|c}
\setlength\tabcolsep{4.5pt}
  & \textbf{Treatment} & \textbf{\rotatebox[origin=c]{90}{SQuirrel}} & \textbf{\rotatebox[origin=c]{90}{JMeter}} & \textbf{\rotatebox[origin=c]{90}{EMF}} & \textbf{\rotatebox[origin=c]{90}{Apache Ant}} & \textbf{\rotatebox[origin=c]{90}{ArgoUML}} & \textbf{\rotatebox[origin=c]{90}{Hibernate}} & \textbf{\rotatebox[origin=c]{90}{JEdit}} & \textbf{\rotatebox[origin=c]{90}{JFreeChart}} & \textbf{\rotatebox[origin=c]{90}{Columba}} & \textbf{\rotatebox[origin=c]{90}{JRuby}} & \textbf{\rotatebox[origin=c]{90}{Median}} & \textbf{\rotatebox[origin=c]{90}{IQR}} \\ 
\hline
& CLA &  65 & \cellcolor[HTML]{FFCCC9}72 & 55 & \cellcolor[HTML]{FFCCC9}67 & 67 & 62 & \cellcolor[HTML]{FFCCC9}79 & \cellcolor[HTML]{FFCCC9}55 & 64 & 65 & 65 & 5 \\ 
& CLAFI+RF  & 53	&	31	&	30	&	46	&	47	&	42	&	39	&	25	&	23	&	23	&	35	&	17 \\ 
 Recall &   CLA+RF	& \cellcolor[HTML]{FFCCC9}81	& \cellcolor[HTML]{FFCCC9}75	& \cellcolor[HTML]{FFCCC9}77	& \cellcolor[HTML]{D28986}91	& \cellcolor[HTML]{FFCCC9}85	& \cellcolor[HTML]{FFCCC9}84	& \cellcolor[HTML]{FFCCC9}76	& \cellcolor[HTML]{D28986}78	& 61	& \cellcolor[HTML]{FFCCC9}82	& 80	& 8 \\ 

$(M=8\%)$ & Easy  & 58	&	\cellcolor[HTML]{FFCCC9}77	&	41	&	27	&	\cellcolor[HTML]{D28986}90	&	\cellcolor[HTML]{FFCCC9}75	&	22	&	\cellcolor[HTML]{FFCCC9} 55	&	\cellcolor[HTML]{FFCCC9}88	&	\cellcolor[HTML]{D28986}90	&	67	&	47 \\ 

& \textbf{Easy+CLA}  & \cellcolor[HTML]{D28986}93	&	\cellcolor[HTML]{D28986}96	&	\cellcolor[HTML]{D28986}82	&	\cellcolor[HTML]{D28986}91	&	\cellcolor[HTML]{D28986}95	&	\cellcolor[HTML]{D28986}95	&	\cellcolor[HTML]{D28986}98	&	\cellcolor[HTML]{D28986}71	&	\cellcolor[HTML]{D28986}98	&	\cellcolor[HTML]{D28986}98	&	95	&	7 \\ 
\hline

\multirow{5}{*}{\rotatebox[origin=c]{0}{\parbox[c]{1cm}{\centering APFD}}} & CLA  & \cellcolor[HTML]{FFCCC9}73 & \cellcolor[HTML]{FFCCC9}73 & \cellcolor[HTML]{FFCCC9}61 & \cellcolor[HTML]{D28986}78 & 72 & 65 & \cellcolor[HTML]{D28986}91 & \cellcolor[HTML]{FFCCC9}61 & 70 & 72 & 72 & 8 \\
& CLAFI+RF  & \cellcolor[HTML]{FFCCC9}74	&	63	&	58	&	64	&	70	&	57	&	76	&	57	&	67	&	68	&	66	&	10 \\ 
 &   CLA+RF	& \cellcolor[HTML]{FFCCC9}74	& \cellcolor[HTML]{FFCCC9}74	& \cellcolor[HTML]{FFCCC9}64	& \cellcolor[HTML]{FFCCC9}73	& 53	& 62	& \cellcolor[HTML]{FFCCC9}80	& \cellcolor[HTML]{FFCCC9}64	& 64	& 68	& 66	& 9 \\ 
$(M=6\%)$ & Easy  & 57	&	\cellcolor[HTML]{FFCCC9}76	&	41	&	27	&	\cellcolor[HTML]{FFCCC9}83	&	\cellcolor[HTML]{FFCCC9}71	&	22	&	54	&	\cellcolor[HTML]{FFCCC9}87	&	\cellcolor[HTML]{FFCCC9}85	&	64	&	42 \\ 
&  \textbf{Easy+CLA}	&	\cellcolor[HTML]{D28986}92	&	\cellcolor[HTML]{D28986}94	&	\cellcolor[HTML]{D28986}81	&	\cellcolor[HTML]{D28986}82	&	\cellcolor[HTML]{D28986}96	&	\cellcolor[HTML]{D28986}93	&	\cellcolor[HTML]{D28986}93	&	\cellcolor[HTML]{D28986}80	&	\cellcolor[HTML]{D28986}98	&	\cellcolor[HTML]{D28986}98	&	93	&	14 \\ 

\hline
\end{tabular}

\vspace{-10pt}
\label{tab:rq1}
\end{table*}

{\bf Cautionary note:}   \citet{Menzies:2007prec} warned that precision can be misleading for imbalanced data sets like those studied here
(e.g. Table~\ref{tab:details} reports that the median of target class is 5\%).  Hence, we will not report precision results and will not place much weight on
 F1.  Table~\ref{tab:rq1} shows the results of the experiment. The Cohen'$d$ effect size test is applied to determine which results are similar by calculating $medium\_{step2}$ or $M$ across Recall and APFD. 

 On each target project in Table \ref{tab:rq1}, the \colorbox[HTML]{D28986}{darker} cells show top rank while the \colorbox[HTML]{FFCCC9}{lighter} cells show rank two. Different colors indicate the statistically significant differences at least equal or larger than the best result(s) subtracts the $medium\_{step2}$ ($M$). Observations from Table~\ref{tab:rq1} include:

\bi
\item Surprisingly, there was no improvements from the data preprocessing (metrics and instances selection) and supervised learning (RF) step after CLA. 

\item CLA performs similarly to Easy in the recall, betters in APFD. In another word, CLA performs better than Easy without access to the train labels. 

\item CLA+RF performs better than Easy in recall and APFD. CLA+RF performs better than Easy without access to the training data's labels. 

\item Notably, CLA+RF performs similarly to CLA, wins in Recall but loses in APFD so CLA's performance is more balanced.

\item Easy+CLA performs the best. This indicates additional effectiveness of CLA as in bettering the performance of Easy. It is notable that Easy relies on training data's labels (i.e., supervised learning) in order to identify SATDs when variants of CLA do not.

\ei 

The results confirm the effectiveness of unsupervised learning (1) by itself or with supervised learning in the case of no labels available, and (2) as a post-processor after the SOTA pattern-based approach. There is a technical-debt proneness tendency within the data to identify SATDs. Moreover, this positive effect also comes with the benefit of not needing access to the labels.

The negative results from preprocessing the data (metrics and instances selection) then supervised learning show that the selected metrics and instances are not fully representative or relevant to identify SATDs on new data. Simply, the technical-debt proneness tendencies that exist inside the training datasets cannot be transferred completely to the target test dataset. Therefore, it might not be effective to learn from the training data at all. 

At the same time, in case there are resources available for labeling the training data, Easy is still a useful pattern-based approach to automatically identify ``easy to find'' SATDs. Then, CLA can be employed to identify the rest of SATDs. The combination of Easy+CLA should be the state-of-the-art automatic method for identifying SATDs. It is simple without any deployment of a machine learning model for the traditional supervised learning approach. 

The findings of this RQ is that:

\begin{RQ}{\normalsize{Result:}} 
From our exploration of various unsupervised learners, the original CLA by Nam et al. performs the best. Moreover, CLA performs similarly to the SOTA pattern-based approach, Easy~\cite{jitterbug}, without having access to the data's labels (100\% less effort).
\end{RQ} \vspace{-15pt}


\textbf{RQ2: How can the state-of-the-art active learning framework be combined with the state-of-the-art unsupervised learner?}



In the situation where business users want to reach a specific amount of SATDs, our previously discussed automatic methods would not be able to guarantee such recall. For this demand, an active learning approach is more suitable~\cite{jitterbug}. However, the SOTA active learner~\cite{jitterbug} requires the training data to be labeled which is very expensive why the RQ1's unsupervised learning method is efficient but not effective enough. Therefore, this RQ explores how to integrate both methods in order to minimize the cost of labeling while improving the effectiveness in identifying SATDs in both settings of training data labels unknown versus known.


\textit{RQ2.1: How to find SATDs with no access to labeled training data?} 

In the low-resource setting (unlabeled training data), unsupervised learning can help guess or pseudo-label the training data with no cost or the experts can label the first 100 instances or 1\% of the test data before applying the active learning strategy. The following methods are tested in this experiment:

\bi
\item \textbf{Emblem}: without utilizing the training data, we start with random 100 test data instances to incrementally update the model to guide the developers to find SATDs.

\item \textbf{Pseudo-labeling} (via CLA) + \textbf{Filtering} (via CLA) + \textbf{Hard}: First, apply unsupervised learning to pseudo-label the training data that highly fits the hypothesized technical-debts proneness tendency with zero human effort. Second, automatically identify the SATDs in test data via the same unsupervised learner. Then, a active learning strategy (as explained in \S\ref{sec:cl_0}) with the data miner RF (referenced in \S\ref{sec:learners}) to
incrementally acquire information and update the model in identifying the SATDs that do not fit such a tendency. 

\item \textbf{Pseudo-labeling} (via CLA) + \textbf{Hard}: This is synonymous with the previous one except skipping the filtering via CLA step and simply apply Hard to incrementally learn and identify SATDs on both frugally pseudo-labeled data and new data. 

\item \textbf{Pseudo-labeling} (via CLA) + \textbf{Filtering} (via CLA) + \textbf{Falcon}: First, apply unsupervised learning to pseudo-label the training data based on the hypothesized technical-debts proneness tendency. Then, the same method is employed to filter out the highly technical-debt prone SATDs in the test data. RF model is applied to continuously train on the pseudo-labeled data and the rest of the test/target data until the found SATDs are 10\% of the estimated SATDs. Finally, the model will discard the cheaply labeled training data while retraining the model on the reviewed test data so far and continuously until the found SATDs are more than the user-specified threshold (i.e. 90\%). 

\item \textbf{Pseudo-labeling} (via CLA) + \textbf{Falcon}: This is synonymous to the previous one except no early filtering via CLA.

\ei

\begin{table*}[!t]
\renewcommand{\arraystretch}{1.1}
\scriptsize
\caption{Comparison between \textbf{Emblem}~\cite{tu2020better}, \textbf{P+Hard}, \textbf{P+F+Hard}, \textbf{P+Falcon}, and \textbf{P+F+Falcon} where \textbf{P} is Pseudo-Labeling and \textbf{F} is Filtering. \textbf{P} and \textbf{F} are done via CLA. The comparison is made in terms of F1 score, G-score, APFD, and cost for identifying SATDs in the low-resource labeled data setting. For F1 score, G-score, and reviewing cost, both methods target at finding 90\% of the SATDs with its estimator. Medians and IQRs (delta between 75th
and 25th percentile, lower the better) are calculated for easy comparisons. Here, the \colorbox[HTML]{FFCCC9}{light red} cells show best performing methods where the difference between them are higher than $M$ reported in the left most column.}
\vspace{-5pt}
\centering
\setlength\tabcolsep{5pt}
\begin{tabular}{c|l|c|c|c|c|c|c|c|c|c|c|c|c}
\setlength\tabcolsep{4.5pt}
  & \textbf{Treatment} & \textbf{\rotatebox[origin=c]{90}{SQuirrel}} & \textbf{\rotatebox[origin=c]{90}{JMeter}} & \textbf{\rotatebox[origin=c]{90}{EMF}} & \textbf{\rotatebox[origin=c]{90}{Apache Ant}} & \textbf{\rotatebox[origin=c]{90}{ArgoUML}} & \textbf{\rotatebox[origin=c]{90}{Hibernate}} & \textbf{\rotatebox[origin=c]{90}{JEdit}} & \textbf{\rotatebox[origin=c]{90}{JFreeChart}} & \textbf{\rotatebox[origin=c]{90}{Columba}} & \textbf{\rotatebox[origin=c]{90}{JRuby}} & \textbf{\rotatebox[origin=c]{90}{Median}} & \textbf{\rotatebox[origin=c]{90}{IQR}} \\
\hline
\multirow{5}{*}{\rotatebox[origin=c]{0}{\parbox[c]{1cm}{\centering F1}}} 
& Emblem\cite{tu2020better} & \cellcolor[HTML]{FFCCC9}61	& 49	& 9	& 8	& \cellcolor[HTML]{FFCCC9}93	& \cellcolor[HTML]{FFCCC9}75	& 7	& 30	& \cellcolor[HTML]{FFCCC9}66	& \cellcolor[HTML]{FFCCC9}81	& 55	& 66 \\ 
& P+Hard	&  28	& 40	& 12	& 17	& 70	& 47	& 32	& 21	& 36	& 52	& 34	& 26 \\
& P+F+Hard & 47 & 59 &  \cellcolor[HTML]{FFCCC9}19 &  \cellcolor[HTML]{FFCCC9}26 &  \cellcolor[HTML]{FFCCC9}88 &  \cellcolor[HTML]{FFCCC9}79 &  \cellcolor[HTML]{FFCCC9}38 &  \cellcolor[HTML]{FFCCC9}39 & 53 &  \cellcolor[HTML]{FFCCC9}86 & 50 & 41 \\
$(M=8\%)$ & \textbf{P+Falcon}	&  50	& \cellcolor[HTML]{FFCCC9}68	& \cellcolor[HTML]{FFCCC9}22	& \cellcolor[HTML]{FFCCC9}29	& \cellcolor[HTML]{FFCCC9}88	& \cellcolor[HTML]{FFCCC9}73	& \cellcolor[HTML]{FFCCC9}41	& \cellcolor[HTML]{FFCCC9}46	& \cellcolor[HTML]{FFCCC9}63	& \cellcolor[HTML]{FFCCC9}83	& 57 &	32 \\
& P+F+Falcon	&  30	& 49	& \cellcolor[HTML]{FFCCC9} 17	&  \cellcolor[HTML]{FFCCC9}22	& 76	& 64	&  \cellcolor[HTML]{FFCCC9}35	&  \cellcolor[HTML]{FFCCC9}39	& 41	& 73	& 40	& 29 \\ 

\hline
\multirow{5}{*}{\rotatebox[origin=c]{0}{\parbox[c]{1cm}{\centering G1}}} 
& Emblem\cite{tu2020better} & \cellcolor[HTML]{FFCCC9}83	& \cellcolor[HTML]{FFCCC9}88	& 20	& 14	& \cellcolor[HTML]{FFCCC9}95	& 82	& 75	& 40	& 82	& 87	& 82	& 47 \\
& P+Hard & 82	& \cellcolor[HTML]{FFCCC9}87	& 75	& \cellcolor[HTML]{FFCCC9}77	& 87	& 77	& \cellcolor[HTML]{FFCCC9}87	& 52	&  \cellcolor[HTML]{FFCCC9}88	& 79	& 81	& 10 \\
& P+F+Hard &	 \cellcolor[HTML]{FFCCC9}83 & 83 &  \cellcolor[HTML]{FFCCC9}87 &  \cellcolor[HTML]{FFCCC9}83 &  \cellcolor[HTML]{FFCCC9}91 &  \cellcolor[HTML]{FFCCC9}86 & 74 &  \cellcolor[HTML]{FFCCC9}78 &  \cellcolor[HTML]{FFCCC9}91 &  \cellcolor[HTML]{FFCCC9}95 & 85 & 8 \\ 
$(M=6\%)$ & \textbf{P+Falcon} & \cellcolor[HTML]{FFCCC9}89	& \cellcolor[HTML]{FFCCC9}90	& 77	& \cellcolor[HTML]{FFCCC9}80	& \cellcolor[HTML]{FFCCC9}94	& \cellcolor[HTML]{FFCCC9}88	& \cellcolor[HTML]{FFCCC9}83	& \cellcolor[HTML]{FFCCC9}74	& \cellcolor[HTML]{FFCCC9}94	& \cellcolor[HTML]{FFCCC9}94	& 89	& 14 \\
& P+F+Falcon & 79	& \cellcolor[HTML]{FFCCC9}86	&75	& 76	& 88	& \cellcolor[HTML]{FFCCC9}85	& 72	& \cellcolor[HTML]{FFCCC9}78	& 84	& \cellcolor[HTML]{FFCCC9}92	& 82	& 10 \\
\hline
\multirow{5}{*}{\rotatebox[origin=c]{0}{\parbox[c]{1cm}{\centering APFD}}} 
& Emblem\cite{tu2020better} &  \cellcolor[HTML]{FFCCC9}92 &  	\cellcolor[HTML]{FFCCC9}92 &  	77 &  	82 &  	\cellcolor[HTML]{FFCCC9}90 &  	 \cellcolor[HTML]{FFCCC9}84 &  	88 &  	\cellcolor[HTML]{FFCCC9}87 &  	\cellcolor[HTML]{FFCCC9}95	 &  \cellcolor[HTML]{FFCCC9}90 &  	89 &  	6  \\

& P+Hard	&  77	&  81	& 66	& 78	& 81	& 67	& 90	& 63	& 81	& 77	& 78	& 14 \\
& P+F+Hard & 86		&  \cellcolor[HTML]{FFCCC9}91		& 79		& \cellcolor[HTML]{FFCCC9}86		&  \cellcolor[HTML]{FFCCC9}89		& 77		& \cellcolor[HTML]{FFCCC9}95		& 75		& 90		& 85		& 86		& 11 \\
$(M=3\%)$ & P+Falcon	&  \cellcolor[HTML]{FFCCC9}90	&  	\cellcolor[HTML]{FFCCC9}93	&  	\cellcolor[HTML]{FFCCC9}81	&  	\cellcolor[HTML]{FFCCC9}84	&  	88	&  	83	&  	\cellcolor[HTML]{FFCCC9}93	&  	69	&  	\cellcolor[HTML]{FFCCC9}93	&  	\cellcolor[HTML]{FFCCC9}89	&  	89	&  	10 \\ 
& \textbf{P+F+Falcon}	&  \cellcolor[HTML]{FFCCC9}91	& \cellcolor[HTML]{FFCCC9}92	& \cellcolor[HTML]{FFCCC9}83	& \cellcolor[HTML]{FFCCC9}86	& \cellcolor[HTML]{FFCCC9}92	& \cellcolor[HTML]{FFCCC9}87	& \cellcolor[HTML]{FFCCC9}93	& 69	& \cellcolor[HTML]{FFCCC9}92	& \cellcolor[HTML]{FFCCC9}91	& 91	& 6 \\

\hline
\multirow{5}{*}{\rotatebox[origin=c]{0}{\parbox[c]{1cm}{\centering Cost}}} 
& \textbf{Emblem}~\cite{tu2020better} & \cellcolor[HTML]{FFCCC9}6	& 13	& \cellcolor[HTML]{FFCCC9}4	& \cellcolor[HTML]{FFCCC9}3	& 17	& \cellcolor[HTML]{FFCCC9}15	& 65	& \cellcolor[HTML]{FFCCC9}4	& \cellcolor[HTML]{FFCCC9}4	& \cellcolor[HTML]{FFCCC9}13	& 10	& 11 \\
& P+Hard	&  20	& 17	& 32	& 25	& 25	& 38	& 10	& 15	& 13	& 24	& 22	& 10 \\
& \textbf{P+F+Hard} & \cellcolor[HTML]{FFCCC9}5 & \cellcolor[HTML]{FFCCC9}3 & 19 & 12 & \cellcolor[HTML]{FFCCC9}11 & \cellcolor[HTML]{FFCCC9}12 & \cellcolor[HTML]{FFCCC9}3 & 10 & \cellcolor[HTML]{FFCCC9}3 & \cellcolor[HTML]{FFCCC9}11 & 11 & 9 \\
$(M=4\%)$  & P+Falcon	&  10	& \cellcolor[HTML]{FFCCC9}7	& 14	& 14	& 19	& 31	& \cellcolor[HTML]{FFCCC9}6	& 11	& \cellcolor[HTML]{FFCCC9}7	& 17	& 13	& 10 \\

& P+F+Falcon &   10	& 9	& 13	& 11	& \cellcolor[HTML]{FFCCC9}13	& 20	& \cellcolor[HTML]{FFCCC9}4	& 12	& 9	& 16	& 12	& 4 \\

\hline
\end{tabular}
\label{tab:rq2}
\vspace{-15pt}
\end{table*}

Beside APFD, we will also evaluate the methods in F1, G1, and the labeling cost. It is recommended to evaluate with metrics that aggregate multiple metrics like F1 and G1. F1 is a harmonic mean of recall and precision, $Precision = TP / (TP + FP)$. G1 is a harmonic mean of recall and false-alarm rate, $FAR = FP / (TP + TN)$:

\vspace{-10pt}

\begin{equation}
\small
F1 = \frac{2 \cdot \mathit{Precision} \cdot \mathit{Recall} }{\mathit{Precision} + \mathit{Recall}}
\end{equation} 

\vspace{-5pt}
\begin{equation}
\small
\mathit{G1} = \frac{2 \cdot \mathit{Recall} \cdot \mathit{(1 - FAR)} }{\mathit{Recall} + (1 - \mathit{FAR})}
\end{equation}

Table \ref{tab:rq2} reports the comparison between all the previously discussed methods across four metrics: F1, G1, APFD, and the labeling Cost. F1, G1, and the labeling cost are measured for all methods aiming to find 90\% of the rest SATDs. Except for the labeling cost, the higher the values of the other metrics the better. The labeling cost is for labeling the test data labels. Pseudo-labeling and Filtering are abbreviated to P and F. Similarly to \textbf{RQ1}, we also employ Cohen'd medium effect size test~\cite{Sawilowsky2009NewES} to determine the best treatment(s) in each target project. From the results, we can see:

\bi
\item \textbf{F1}: P+Falcon performs the best (9/10) while P+F+Hard performs the second-best (7/10). 

\item \textbf{G1}: P+Falcon outperforms the rest (9/10) while P+F+Hard's performance is placed at second.

\item \textbf{APFD}: P+F+Falcon's performance is the best (9/10) while P+Falcon and Emblem perform similarly as the second best (7/10). 

\item \textbf{Cost}: Emblem's and P+F+Hard's processes are the most efficient (7/10). 
\ei

P+F+Hard always performs better than P+Hard across all four metrics. However, the same effect is not observed in P+Falcon and P+F+Falcon due to Falcon losing some insights of frugally pseudo-labeled data after dropping the training data at a certain threshold and the filtering step further reduce the insights (i.e, highly technical-debt prone comments) for Falcon to learn. However, P+Falcon outperforms all methods, except slightly lost to P+F+Falcon in G1, and makes the system experts read 3\% more on average. Falcon's effectiveness overwrites the necessity of filtering. Moreover, P+Falcon's performance is better than P+Hard indicating the effectiveness of Falcon over Hard. 

\begin{figure*}[!b]
\vspace{-7pt}
\begin{center}

\includegraphics[width=\textwidth]{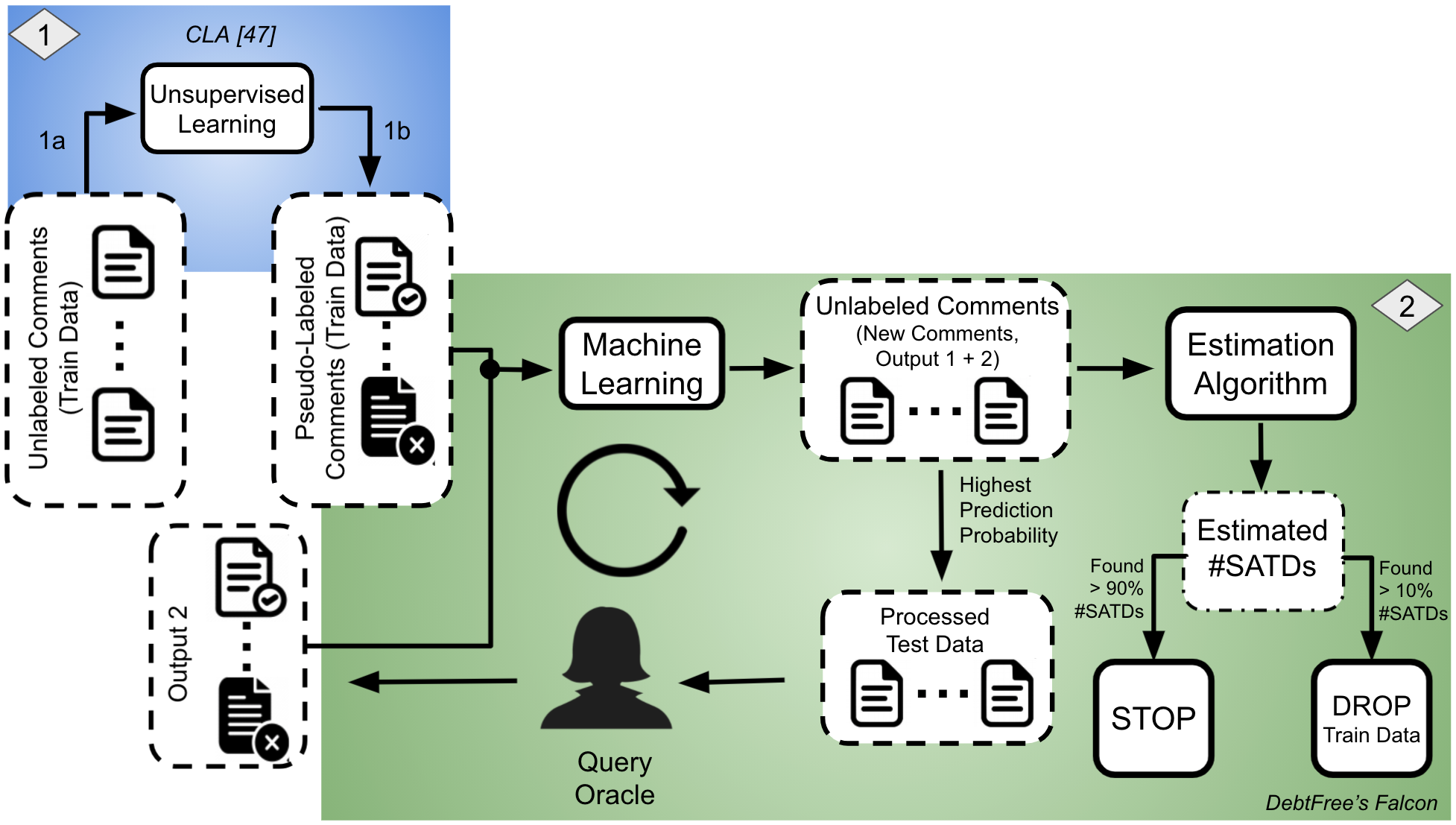}
\vspace{-18pt}
\caption{Workflows of \IT{}(0) = Pseudo-Labeling (via Unsupervised Learning, i.e., CLA~\cite{nam2015clami}) + Active Learning (via Falcon).}
\label{fig:workflow_debtfree0}

\end{center}
\vspace{-25pt}
\end{figure*}

Overall, in the situation of having no access to the labeled training data, the effectiveness of Pseudo-labeling (via CLA) + Falcon demonstrates that:

\bi

\item Unsupervised learning can cheaply and quickly guess the labels of the training data to provide insights for bootstrapping the active learning strategy. However, those frugally pseudo-labeled data do not have enough insights to help Hard guide the experts efficiently to identify the SATDs the whole way up to 90\%.

\item Emblem alone is cost-efficient but it does not effectively identify the SATDs. 

\item A hybrid of both is more preferable as Falcon would drop training data with guessed labels after the found SATDs are at least 10\% of the estimated SATD\% so there are enough insights to help Falcon guide the experts to efficiently identify the rest of SATDs. 

\item The best configuration for a ``label-free'' process including a labeler to utilize the unlabeled data for bootstrapping and the tuned active learning strategy is labeling via CLA and Falcon. Hence, \IT{}(0) = \textbf{Pseudo-labeling} (via CLA) + \textbf{Falcon} as shown in Figure~\ref{fig:workflow_debtfree0}. 

\ei


\textit{RQ2.2: How to find SATDs with having access to labeled training data?}

For this research question, we are interested in the best configurations of \IT{}(100) to take advantage of an abundant amount of labeled data from previous research work~\cite{maldonado2015detecting, jitterbug}. The comparison is made through the F1 score, G1 score, APFD, and cost metrics. The methods are similar to the previous RQ but without the labeling step:

\bi

\item \textbf{Filtering} (via CLA) + \textbf{Hard}: First, apply unsupervised learning to learn the hypothesized technical-debts proneness tendency on the humanly labeled training data then automatically identify the SATDs in test data with zero human effort. Second, an active learning strategy adopted from the SOTA work~\cite{jitterbug} (\S\ref{sec:continuous_learning}) with the data miner RF (\S\ref{sec:learners}) to
incrementally acquire information and update the model in identifying the SATDs that do not fit such tendency. 
\item \textbf{Hard} (\S\ref{sec:continuous_learning}): This is synonymous with the previous one except no filtering via CLA in the beginning. 

\item \textbf{Filtering} (via CLA) + \textbf{Falcon}: The filtering step is the same as above. Then, an RF model is applied to continuously learn on the unfiltered training data and the rest of the unfiltered test/target data until the found SATDs are 10\% of the estimated SATDs. Finally, the model will discard the labeled training data while retraining the model on the reviewed test data so far and continuously until the found SATDs are more than the user-specified threshold (i.e. 90\%). 

\item \textbf{Falcon} (\S\ref{sec:cl_0}): This method is synonymous with the previous one except for no filtering step in the beginning.

\begin{table*}[!t]
\renewcommand{\arraystretch}{1.1}
\scriptsize
\caption{Comparison between \textbf{Hard}, \textbf{F+Hard}, \textbf{Falcon}, and \textbf{F+Falcon} in terms of F1 score, G-score, APFD, and cost in identifying SATDs in the high-resource labeled data setting. Filtering-\textbf{F} are done via CLA. For F1 score, G-score, and reviewing cost, both methods target at finding 90\% of the ``hard to find'' SATDs with its estimator. Medians and IQRs (delta between 75th
and 25th percentile, lower the better) are calculated for easy comparisons. Here, the \colorbox[HTML]{FFCCC9}{light red} cells show best performing methods where the difference between them are higher than $M$ reported in the left most column.}
\vspace{-5pt}
\centering
\setlength\tabcolsep{5pt}
\begin{tabular}{c|l|c|c|c|c|c|c|c|c|c|c|c|c}
\setlength\tabcolsep{4.5pt}
  & \textbf{Treatment} & \textbf{\rotatebox[origin=c]{90}{SQuirrel}} & \textbf{\rotatebox[origin=c]{90}{JMeter}} & \textbf{\rotatebox[origin=c]{90}{EMF}} & \textbf{\rotatebox[origin=c]{90}{Apache Ant}} & \textbf{\rotatebox[origin=c]{90}{ArgoUML}} & \textbf{\rotatebox[origin=c]{90}{Hibernate}} & \textbf{\rotatebox[origin=c]{90}{JEdit}} & \textbf{\rotatebox[origin=c]{90}{JFreeChart}} & \textbf{\rotatebox[origin=c]{90}{Columba}} & \textbf{\rotatebox[origin=c]{90}{JRuby}} & \textbf{\rotatebox[origin=c]{90}{Median}} & \textbf{\rotatebox[origin=c]{90}{IQR}} \\
\hline
\multirow{3}{*}{\rotatebox[origin=c]{0}{\parbox[c]{1cm}{\centering F1}}} 
& \textbf{Hard}	&  \cellcolor[HTML]{FFCCC9}48 & 59 & \cellcolor[HTML]{FFCCC9}28 & 25 & \cellcolor[HTML]{FFCCC9}94 & \cellcolor[HTML]{FFCCC9}82 & 29 & \cellcolor[HTML]{FFCCC9}53 & \cellcolor[HTML]{FFCCC9}84 & \cellcolor[HTML]{FFCCC9}92 & 56 & 55 \\
& F+Hard & 34 & 52 & \cellcolor[HTML]{FFCCC9}23 & 24 & 82 & 66 & \cellcolor[HTML]{FFCCC9}44 & \cellcolor[HTML]{FFCCC9}51 & 73 & 80 & 52 & 29 \\
$(M=8\%)$  & \textbf{Falcon} &   \cellcolor[HTML]{FFCCC9}55 & \cellcolor[HTML]{FFCCC9}70 & \cellcolor[HTML]{FFCCC9}28 & \cellcolor[HTML]{FFCCC9}44 & \cellcolor[HTML]{FFCCC9}93 & 73 & \cellcolor[HTML]{FFCCC9}50 & \cellcolor[HTML]{FFCCC9}48 & 54 & 82 & 55 & 25 \\
& F+Falcon &  34 & 45 & 18 & 24 & 84 & 66 & 30 & 34 & 35 & 70 & 35 & 36 \\ 

\hline
\multirow{3}{*}{\rotatebox[origin=c]{0}{\parbox[c]{1cm}{\centering G1}}} 
& \textbf{Hard} & \cellcolor[HTML]{FFCCC9}93 & \cellcolor[HTML]{FFCCC9}92 & \cellcolor[HTML]{FFCCC9}88 & \cellcolor[HTML]{FFCCC9}87 & \cellcolor[HTML]{FFCCC9}98 & \cellcolor[HTML]{FFCCC9}89 & \cellcolor[HTML]{FFCCC9}93 & \cellcolor[HTML]{FFCCC9}82 & \cellcolor[HTML]{FFCCC9}97 & \cellcolor[HTML]{FFCCC9}96 & 93 & 8  \\
& F+Hard &  90 & \cellcolor[HTML]{FFCCC9}91 & \cellcolor[HTML]{FFCCC9}86 & \cellcolor[HTML]{FFCCC9}85 & 95 & 86 & \cellcolor[HTML]{FFCCC9}92 & \cellcolor[HTML]{FFCCC9}80 & \cellcolor[HTML]{FFCCC9}98 & \cellcolor[HTML]{FFCCC9}94 & 91 & 8  \\ 
$(M=2\%)$  & Falcon & 86 & \cellcolor[HTML]{FFCCC9}90 & 83 & 78 & \cellcolor[HTML]{FFCCC9}96 & \cellcolor[HTML]{FFCCC9}90 & 88 & \cellcolor[HTML]{FFCCC9}80 & \cellcolor[HTML]{FFCCC9}96 & \cellcolor[HTML]{FFCCC9}94 & 89 & 11 \\
& F+Falcon & 88 & 89 & 81 & 82 & 94 & \cellcolor[HTML]{FFCCC9}88 & \cellcolor[HTML]{FFCCC9}91 & 78 & 93 & 92 & 89 & 10 \\
\hline
\multirow{3}{*}{\rotatebox[origin=c]{0}{\parbox[c]{1cm}{\centering APFD}}} 

& Hard &  95 & \cellcolor[HTML]{FFCCC9}95 & \cellcolor[HTML]{FFCCC9}95 & 91 & 91 & 89 & 95 & \cellcolor[HTML]{FFCCC9}90 & \cellcolor[HTML]{FFCCC9}98 & 92 & 94 & 4 \\
& \textbf{F+Hard} & \cellcolor[HTML]{FFCCC9}98 & \cellcolor[HTML]{FFCCC9}96 & \cellcolor[HTML]{FFCCC9}97 & \cellcolor[HTML]{FFCCC9}96 & \cellcolor[HTML]{FFCCC9}97 & \cellcolor[HTML]{FFCCC9}94 & \cellcolor[HTML]{FFCCC9}96 & \cellcolor[HTML]{FFCCC9}89 & \cellcolor[HTML]{FFCCC9}98 & \cellcolor[HTML]{FFCCC9}96 & 96 & 1 \\
$(M=2\%)$ & Falcon &  94 & \cellcolor[HTML]{FFCCC9}95 & 91 & 90 & 91 & 89 & 94 & 83 & \cellcolor[HTML]{FFCCC9}97 & 92 & 92 & 4 \\ 
& F+Falcon &  \cellcolor[HTML]{FFCCC9}97 & \cellcolor[HTML]{FFCCC9}97 & 94 & 92 & \cellcolor[HTML]{FFCCC9}96 & \cellcolor[HTML]{FFCCC9}92 & \cellcolor[HTML]{FFCCC9}98 & 76 & \cellcolor[HTML]{FFCCC9}98 & \cellcolor[HTML]{FFCCC9}96 & 96 & 5 \\

\hline
\multirow{3}{*}{\rotatebox[origin=c]{0}{\parbox[c]{1cm}{\centering Cost}}} 
& Hard & 13 & 11 & 14 & 20 & 18 & 17 & 14 & 10 & \cellcolor[HTML]{FFCCC9}4 & 14 & 14 & 6 \\
& \textbf{F+Hard} & \cellcolor[HTML]{FFCCC9}8 & \cellcolor[HTML]{FFCCC9}7 & \cellcolor[HTML]{FFCCC9}5 & 12 &  \cellcolor[HTML]{FFCCC9}11  & \cellcolor[HTML]{FFCCC9}10  & 9  & \cellcolor[HTML]{FFCCC9}7  & \cellcolor[HTML]{FFCCC9}5  & \cellcolor[HTML]{FFCCC9}10 & 9 & 3 \\
$(M=2\%)$ & Falcon &   \cellcolor[HTML]{FFCCC9}9 & \cellcolor[HTML]{FFCCC9}7 & 11 & \cellcolor[HTML]{FFCCC9}9 & \cellcolor[HTML]{FFCCC9}12 & 20 & \cellcolor[HTML]{FFCCC9}3 & \cellcolor[HTML]{FFCCC9}9 & 8 & 14 & 9 & 4  \\

& F+Falcon &   \cellcolor[HTML]{FFCCC9}8 & \cellcolor[HTML]{FFCCC9}7 & 12 & \cellcolor[HTML]{FFCCC9}7 & 17 & 24 & 6 & 11 & 9 & 17 & 10 & 10 \\

\hline
\end{tabular}
\label{tab:rq3}
\vspace{-15pt}
\end{table*}

\ei


The results for this RQ are reported in Table \ref{tab:rq3}. Similarly, F1, G1, and the labeling cost are measured for all methods aiming to find 90\% of the rest SATDs. Except for the labeling cost, the higher the values of the other metrics the better. The labeling cost is for labeling the test data labels. Filtering is abbreviated to F. Similarly to \textbf{RQ1}, we also employ Cohen'd medium effect size test \cite{Cohen:1995,Sawilowsky2009NewES} to determine the best treatment to determine the best treatments in each target project. From the results, it is observed that:

\bi

\item \textbf{F1}: Hard's and Falcon's performance are similarly as the best ones (7/10). 

\item \textbf{G1}: Hard outshines the rest over all the projects while F+Hard performs as the second best (7/10).


\item \textbf{APFD}: F+Hard outperforms the rest on ten out of ten projects and F+Falcon comes in as second place (7/10). 


\item \textbf{Cost}: F+Hard is the most efficient in eight out of ten projects and then Falcon (6/10). 

\ei

\begin{figure*}[!t]
\vspace{-7pt}
\begin{center}

\includegraphics[width=\textwidth]{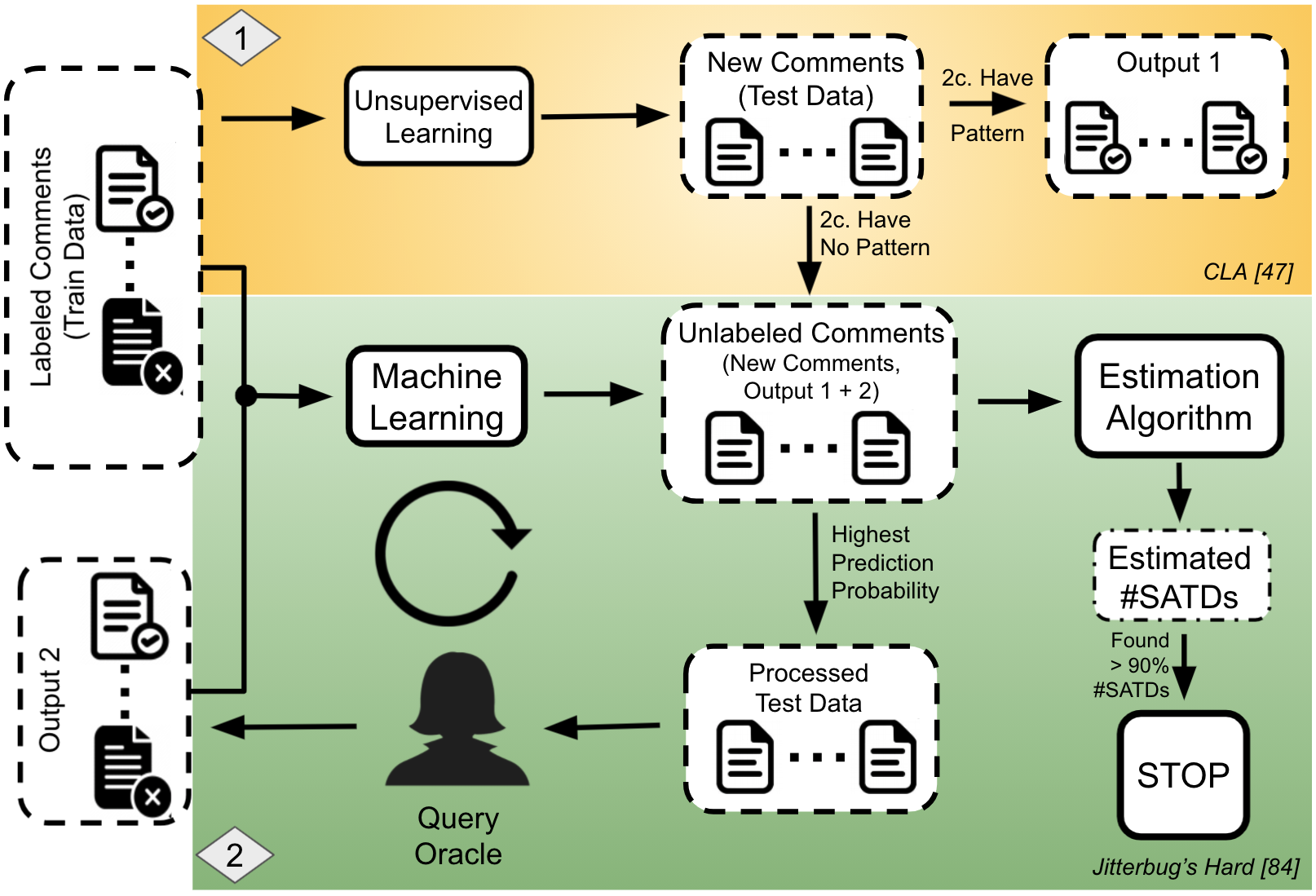}
\vspace{-18pt}
\caption{Workflows of \IT{}(100) = Filtering (via Unsupervised Learning, i.e., CLA~\cite{nam2015clami}) + Active Learning (via Hard~\cite{jitterbug}). }
\label{fig:workflow_debtfree100}

\end{center}
\vspace{-25pt}
\end{figure*}

Falcon still outperforms F+Falcon across F1, G1, and cost while losing on APFD. However, Hard outshines Falcon here, wins in cost, draws in F1, and loses in G1 and APFD. This indicates that when there is an abundant amount of labeled data, discarding them after reaching a certain threshold is not as effective as just active learning on the rest since there is enough insights from the labels. Hard performs similarly to F+Hard as the best ones across all methods by winning on F1 and G1 while losing on APFD and cost. However, the benefit of filtering here is it helps stabilize Hard's performance as Hard by itself has a high variance across four metrics. As mentioned previously RQ1, F1 and G1 scores are only single points of the curve while APFD calculates the area under the recall-cost curve. In the theme of effort-aware, the best configurations of \IT{}(100) include \textbf{Filtering} (via CLA) with \textbf{Hard} as shown in Figure~\ref{fig:workflow_debtfree100}. 

From this RQ, we conclude that: 

\begin{RQ}{\normalsize{Result:}} 
With the effort-aware theme, we investigate different combinations of active learners and CLA across two settings of the training data, either with having  1-\textit{no access} and 2-\textit{access} to the labels to propose \IT{}. In setting 1, or \IT{}(0), the best combination is Pseudo-Labeling (via CLA) with our proposed active learner, Falcon. In setting 2, or \IT{}(100), the best combination is Filtering (via CLA) with the SOTA active learner for SATDs identification, Hard.
\end{RQ} \vspace{-15pt}

\textbf{RQ3: How does the proposed \IT{} perform against state-of-the-art models in identifying SATDs?}

For this research question, we are interested in the overall performance of \IT{} by comparing (1) ``label-free'' \IT{}(0), (2) the latest state-of-the-art deep convolutional neural network-based approach \cite{ren2019neural}, (3) two-step SOTA Jitterbug approach \cite{jitterbug}, and (4) \IT{}(100). The comparison is made via F1 score, G1 score, APFD, and cost metrics:

\bi
\item \textbf{DebtFree(0)}: First, apply unsupervised learning to label the unlabeled training data based on the hypothesized technical-debts proneness tendency. Finally, Falcon is applied to continuously learn and identify the rest of SATDs.

\item \textbf{Jitterbug~\cite{jitterbug}}: First apply pattern-based approach (i.e, Easy, described in \S5's RQ1) to automatically identify the ``easy
to find'' SATDs, then apply a active learning strategy (i.e., Hard, described in \S\ref{sec:continuous_learning}) with RF as the learner of choice to
guide humans in identifying the ``hard to find'' SATDs.

\item \textbf{CNN}: deep learning solution that is based on a convolutional neural network structure with word2vec features and hyperparameter tuning to classify each comment
into SATD or non-SATD \cite{ren2019neural}. 

\item \IT{}(100): First, apply CLA as an unsupervised learner to filter out the test data that are most likely to be technical debts. Then, the SOTA active learning strategy of Hard is applied to incrementally identify the rest SATDs. 

\ei


\begin{table*}[!t]
\renewcommand{\arraystretch}{1.1}
\scriptsize
\caption{Comparison between \IT(0), \textbf{CNN}, \textbf{Jitterbug}, and \IT(100) in terms of F1 score, G-score, APFD, and cost. For F1 score, G-score, and reviewing cost, both methods target at finding 90\% of the ``hard to find'' SATDs with its estimator. Medians and IQRs (delta between 75th
and 25th percentile, lower the better) are calculated for easy comparisons.  Here,  the \colorbox[HTML]{D28986}{darker} cells show top rank while the \colorbox[HTML]{FFCCC9}{lighter} cells show rank two where the difference between two ranks is statistically significant by being higher than $M$ reported in the left most column.}
\centering
\setlength\tabcolsep{5pt}
\begin{tabular}{l|l|c|c|c|c|c|c|c|c|c|c|c|c}
\setlength\tabcolsep{4.5pt}
  & \textbf{Treatment} & \textbf{\rotatebox[origin=c]{90}{SQuirrel}} & \textbf{\rotatebox[origin=c]{90}{JMeter}} & \textbf{\rotatebox[origin=c]{90}{EMF}} & \textbf{\rotatebox[origin=c]{90}{Apache Ant}} & \textbf{\rotatebox[origin=c]{90}{ArgoUML}} & \textbf{\rotatebox[origin=c]{90}{Hibernate}} & \textbf{\rotatebox[origin=c]{90}{JEdit}} & \textbf{\rotatebox[origin=c]{90}{JFreeChart}} & \textbf{\rotatebox[origin=c]{90}{Columba}} & \textbf{\rotatebox[origin=c]{90}{JRuby}} & \textbf{\rotatebox[origin=c]{90}{Median}} & \textbf{\rotatebox[origin=c]{90}{IQR}} \\
\hline
\multirow{3}{*}{\rotatebox[origin=c]{0}{\parbox[c]{1cm}{\centering F1}}} 
& DebtFree(0)	&  \cellcolor[HTML]{FFCCC9}50	& \cellcolor[HTML]{D28986}68	& 22	& \cellcolor[HTML]{FFCCC9}29	&  \cellcolor[HTML]{D28986}88	& \cellcolor[HTML]{D28986}73	& \cellcolor[HTML]{FFCCC9}41	& 46	& \cellcolor[HTML]{FFCCC9}63	& \cellcolor[HTML]{D28986}83	& 57 &	32 \\
& CNN~\cite{ren2019neural} & \cellcolor[HTML]{D28986}60	& \cellcolor[HTML]{D28986}62	& \cellcolor[HTML]{FFCCC9}51	& \cellcolor[HTML]{D28986}45	& \cellcolor[HTML]{D28986}82	& \cellcolor[HTML]{D28986}78	& \cellcolor[HTML]{D28986}52	& \cellcolor[HTML]{FFCCC9}51	& \cellcolor[HTML]{D28986}76	& \cellcolor[HTML]{D28986}87	& 61	& 27 \\ 
$(M=8\%)$ & Jitterbug~\cite{jitterbug}  & 24 & 55 & \cellcolor[HTML]{D28986}61 & 17 & 21 & 34 & 23 & \cellcolor[HTML]{D28986}75 & 47 & \cellcolor[HTML]{FFCCC9}23 & 29 & 32 \\
& \textbf{DebtFree(100)}    & 34	  & \cellcolor[HTML]{FFCCC9}52	  & 23	  & \cellcolor[HTML]{FFCCC9}24	  & \cellcolor[HTML]{D28986}82	  & \cellcolor[HTML]{FFCCC9}66	  & \cellcolor[HTML]{FFCCC9}44	  & 51	  & \cellcolor[HTML]{D28986}73	  & \cellcolor[HTML]{D28986}80	  & 52	  & 29 \\
\hline
\multirow{3}{*}{\rotatebox[origin=c]{0}{\parbox[c]{1cm}{\centering G1}}} 
& DebtFree(0) &\cellcolor[HTML]{D28986}89	& \cellcolor[HTML]{D28986}90	& 77	& \cellcolor[HTML]{FFCCC9}80	& \cellcolor[HTML]{D28986}94	& \cellcolor[HTML]{D28986}88	& \cellcolor[HTML]{FFCCC9}83	& 74	& \cellcolor[HTML]{FFCCC9}94	& \cellcolor[HTML]{D28986}94	& 89	& 14 \\
& CNN~\cite{ren2019neural} & \cellcolor[HTML]{FFCCC9}87	& \cellcolor[HTML]{D28986}91	& \cellcolor[HTML]{FFCCC9}83	& \cellcolor[HTML]{D28986}84	& \cellcolor[HTML]{FFCCC9}92	& \cellcolor[HTML]{D28986}90	& 79	& 76	& \cellcolor[HTML]{D28986}98	& \cellcolor[HTML]{D28986}95	& 89	& 33 \\ 
$(M=2\%)$ & \textbf{Jitterbug}~\cite{jitterbug} &  \cellcolor[HTML]{D28986}91	&	\cellcolor[HTML]{FFCCC9}85	&		\cellcolor[HTML]{D28986}88	&		\cellcolor[HTML]{D28986}85	&		 \cellcolor[HTML]{D28986}93	&		\cellcolor[HTML]{D28986}88	&		\cellcolor[HTML]{D28986}91	&		\cellcolor[HTML]{D28986}93	&	90	&		\cellcolor[HTML]{FFCCC9}88	&		89	&		3\\
& \textbf{DebtFree(100)}  & 	\cellcolor[HTML]{D28986}90 & 			\cellcolor[HTML]{D28986}91 & 			\cellcolor[HTML]{D28986}86 & 			\cellcolor[HTML]{D28986}85 & 			\cellcolor[HTML]{D28986}95 & 			\cellcolor[HTML]{D28986}86 & 			\cellcolor[HTML]{D28986}92 & 			\cellcolor[HTML]{FFCCC9}80 & 			\cellcolor[HTML]{D28986}98 & 			\cellcolor[HTML]{D28986}94 & 			90 & 			8\\
\hline
\multirow{3}{*}{\rotatebox[origin=c]{0}{\parbox[c]{1cm}{\centering APFD}}} 
& DebtFree(0)	&  \cellcolor[HTML]{FFCCC9}90	&  	\cellcolor[HTML]{FFCCC9}93	&  	\cellcolor[HTML]{FFCCC9}81	&  \cellcolor[HTML]{FFCCC9}	84	&  	88	&  	\cellcolor[HTML]{FFCCC9}83	&  	\cellcolor[HTML]{FFCCC9}93	&  	69	&  	\cellcolor[HTML]{FFCCC9}93	&  	\cellcolor[HTML]{FFCCC9}89	&  	89	&  	10 \\ 
& CNN~\cite{ren2019neural} & 69	& 83	& 58	& 66	& 87	& 79	& 64	& 81	& 89	& \cellcolor[HTML]{FFCCC9}88	& 80	& 21 \\ 
$(M=4\%)$ & \textbf{Jitterbug}~\cite{jitterbug}  & 	\cellcolor[HTML]{D28986}95 & 		\cellcolor[HTML]{D28986}99 & 	\cellcolor[HTML]{D28986}95 & 	\cellcolor[HTML]{D28986}93 & 	\cellcolor[HTML]{FFCCC9}92 & 	\cellcolor[HTML]{D28986}94 & 	\cellcolor[HTML]{D28986}100 & 	\cellcolor[HTML]{D28986}99 & 	\cellcolor[HTML]{D28986}95 & 	\cellcolor[HTML]{D28986}94 & 	95 & 	5\\
& \textbf{DebtFree(100)}  &  	\cellcolor[HTML]{D28986}98 & 		\cellcolor[HTML]{D28986}96 & 		\cellcolor[HTML]{D28986}97 & 		\cellcolor[HTML]{D28986}96 & 		\cellcolor[HTML]{D28986}97 & 		\cellcolor[HTML]{D28986}94 & 	\cellcolor[HTML]{D28986}	96 & 		\cellcolor[HTML]{FFCCC9}89 & 		\cellcolor[HTML]{D28986}98 & 		\cellcolor[HTML]{D28986}96 & 		96 & 		1\\

\hline
\multirow{3}{*}{\rotatebox[origin=c]{0}{\parbox[c]{1cm}{\centering Cost}}} 
& DebtFree(0)	&  \cellcolor[HTML]{FFCCC9}10	& \cellcolor[HTML]{FFCCC9}7	& 14	& \cellcolor[HTML]{FFCCC9}14	& 19	& 31	& \cellcolor[HTML]{FFCCC9}6	& 11	& \cellcolor[HTML]{FFCCC9}7	& 17	& 13	& 10 \\
& \textbf{CNN}~\cite{ren2019neural} & \cellcolor[HTML]{D28986}0	& \cellcolor[HTML]{D28986}0	& \cellcolor[HTML]{D28986}0	& \cellcolor[HTML]{D28986}0	& \cellcolor[HTML]{D28986}0	& \cellcolor[HTML]{D28986}0	& \cellcolor[HTML]{D28986}0	& \cellcolor[HTML]{D28986}0	& \cellcolor[HTML]{D28986}0	& \cellcolor[HTML]{D28986}0	& 0	& 0 \\ 
$(M=4\%)$ & Jitterbug~\cite{jitterbug}  & 18 &  35 & 28 & 24 & \cellcolor[HTML]{FFCCC9}12 & 25 & 17 & 19 & 23 & 27 & 24 & 8 \\
 & \textbf{DebtFree(100)}  & \cellcolor[HTML]{FFCCC9}8 & \cellcolor[HTML]{FFCCC9}7 & \cellcolor[HTML]{FFCCC9}5 & \cellcolor[HTML]{FFCCC9}12 & \cellcolor[HTML]{FFCCC9}11 & \cellcolor[HTML]{FFCCC9}10 & 9 & \cellcolor[HTML]{FFCCC9}7 & \cellcolor[HTML]{FFCCC9}5 & \cellcolor[HTML]{FFCCC9}10 & 9 & 3\\

\hline
\end{tabular}
\label{tab:rq4}
\vspace{-15pt}
\end{table*}

The results for this RQ are reported in Table \ref{tab:rq4}. Similarly, F1, G1, and the labeling cost are measured for all methods aiming to find 90\% of the rest SATDs. Except for the labeling cost, the higher the values of the other metrics the better. The labeling cost is for labeling the test data labels. Similar to \textbf{RQ1}, we also employ Cohen'd medium effect size test \cite{Cohen:1995} to determine the best treatment to determine the best treatments in each target project. From the results, it is observed that:

\bi

\item \textbf{F1}: CNN outperforms the rest of the methods in eight out of ten projects and then \IT{}(0)'s performance is placed at second. 


\item \textbf{G1}: \IT{}(100) outshines the rest in nine projects and then Jitterbug comes in second place.


\item \textbf{APFD}: \IT{}(100) performs similarly as Jitterbug as the best ones in nine out of ten projects. 


\item \textbf{Cost}: CNN is the most efficient since it is modeled as the traditional supervised approach (i.e., not touching the test/target data).

\ei





\IT{}(100) performs similarly to the SOTA supervised learning approach of CNN \cite{ren2019neural}, wins in APFD, draws in G1, and loses in F1 and cost. However, \IT{}(100)'s performance exceeds the SOTA two-step Jitterbug approach \cite{jitterbug} across three metrics (F1, G1, and cost) while performing similarly in APFD. In particular, \IT{}(100) saves the labeling cost 250\% from Jitterbug on average (except in \textit{ArgoUML} projects). 

\IT{}(0) outperforms the SOTA two-step Jitterbug approach \cite{jitterbug} while performing similarly to the SOTA supervised learning approach of CNN \cite{ren2019neural}. \IT{}(0) wins CNN in APFD, draws in G1, and loses in F1 and cost. \IT{}(0) wins Jitterbug in F1 and cost, draws in G1, and loses in APFD. Specifically, \IT{}(0) saves the labeling cost almost double than Jitterbug on average (except in \textit{ArgoUML} and \textit{Hibernate} project). However, \IT{}(0) loses to \IT{}(100) across four metrics. In general, both of these SOTA work's and \IT{}(100)'s effectiveness rely on the labeled training data, without them, they might not exist. On the other hand, \IT{}(0) does not require the training data to be labeled, and requiring no labels from the training data is already a win in itself. Specifically, out of the total ten datasets, there are nine labeled training datasets ($90\%$) so CNN needs 90\% labeled data while Jitterbug needs 92\% on average while \textit{\IT{}(0) only need 1\% of the data to be labeled (99\% cheaper)}.

Notably, Falcon having access to labeled training data actually surpasses CNN on three metrics (F1, G1, and APFD) while losing in cost. Moreover, Falcon performs better than Jitterbug in F1 and cost, draws in G1, and loses in APFD. Specifically, Falcon saves the labeling cost 240\% from Jitterbug on average (except in \textit{ArgoUML} project). Hence, \textbf{Falcon} is more balanced than \textbf{Filtering} (via CLA) and \textbf{Hard} in term of effectiveness over both methods.

On a side note, active learning without the training data, Emblem, performs similarly to Jitterbug (with access to labeled training data and a pattern-based approach) as Emblem wins in F1 and Cost but loses in G1 and APFD. Moreover, Emblem loses to Jitterbug by a small margin for G1 (median at 82 versus 89) versus APFD (median at 89 versus 95) but wins over by a larger margin for F1 (median at 55 versus 29) and Cost (median at 13 versus 24). This restates how labeled training data and insights from them are not completely transferred to the learning of the continuous strategy which contradicts the previous SOTA's conclusion. 

Altogether, \IT{}'s effectiveness can be concluded that:

\begin{RQ}{\normalsize{Result:}} 
When comparing  against the SOTA semi-supervised learning work by \citet{jitterbug} and the SOTA supervised learning work (with deep learning) by \citet{ren2019neural}, our proposed method \IT{} outperforms them significantly. First, \IT{}(100) performs similarly to \citet{ren2019neural}'s work and better than \citet{jitterbug}'s work while reducing the labeling cost by 2.5 times. Second, \IT{}(0) performs similarly to \citet{ren2019neural}'s work without having access to the training data's labels and outperforms \citet{jitterbug}'s work while being 99\% less effort.
\end{RQ} \vspace{-15pt}

\begin{table*}[!t]
\scriptsize
\caption{The left subtable showed TP/P rate after Easy and CLA while the right subtable are the median ratios of  estimated SATDs over actual SATDs per iteration of the active learning strategies of Emblem, \IT{}(0)'s-D(0)'s Falcon, \IT{}(100)'s-D(100)'s Hard, and Jitterbug's-J's Hard ~\cite{jitterbug}. The median and IQR are also reported for ease of comparison.}
\vspace{-5pt}
\resizebox{\linewidth}{!}{\minipage{0.4\linewidth}%
\centering
\setlength\tabcolsep{2pt}
\subcaption{Percentage of SATDs being identified by the Easy and CLA in each project.}
\begin{tabular}{l|c|c}
\setlength\tabcolsep{4.5pt} 
\textbf{Datasets} & Easy    & CLA  \\ \hline
\textbf{SQuirrel} 	&	58  & 47	 \\
\textbf{JMeter} 	&	77 & 22	 \\
\textbf{EMF} 	&	41 & 25	 \\
\textbf{Apache Ant} 	&	27 & 42	 \\
\textbf{ArgoUML} 	&	90 & 45	 \\
\textbf{Hibernate} 	&	76 & 31	 \\
\textbf{JEdit} 	&	22 & 35	 \\
\textbf{JFreeChart} 	&	55 & 13	 \\
 \textbf{Columba} 	&	88 & 18	 \\ 
\textbf{JRuby} 	&	90 &  27	 \\ \hline
\textbf{Median} 	&	67 & 29	 \\
\textbf{IQR} 	&	47 	& 13  \\ \hline
\end{tabular} 
\endminipage
\hspace{20pt}\minipage{0.5\linewidth}%
\centering
 \subcaption{Median ratio of estimated SATDs over actual SATDs across all iterations per project by each active learning method. The higher the number than 100\%, the more the approach overestimate and vice-versa. }
\setlength\tabcolsep{2pt}
\begin{tabular}{l|c|c|c|c}
\setlength\tabcolsep{4.5pt} 
\textbf{Datasets} & Emblem & D(0)'s & D(100)'s & J's \\ 
 &  & Falcon  & Hard & Hard \\ \hline 
\textbf{SQuirrel} 	&	81 	&		88 	&		69 	&		153 \\
\textbf{JMeter} 	&	 94 	&		104 	&		83	 	&	429 \\
\textbf{EMF} &	13 	&		101 	&		75 	&		76 \\
\textbf{Apache Ant} 	& 7 	&		60 	&		100 	&		23 \\
\textbf{ArgoUML} 	&	125 	&		93 	&		58	 	&	1195 \\
\textbf{Hibernate} 	&	74 	&		97 	&		85	 	&	362 \\
\textbf{JEdit} 	&	122 	&		71  	&		85 	&		148 \\
\textbf{JFreeChart} 	&	34 	&		60 	&		81 	&		128 \\
 \textbf{Columba} 	& 60 	&		98 	&		86 	&		592 \\ 
\textbf{JRuby} 	&	72 	&		103 	&		84 	&		826 \\ \hline
\textbf{Median} 	&	73 	&		95 	&		84 	&		258 \\
\textbf{IQR} 	&	60 	&		30 	&		4 	&		464 \\ \hline
\end{tabular}
\endminipage}

\label{tab:rq5}
\end{table*}

\section{Threats of Validity}\label{tion:threats}

There are several validity threats~\cite{feldt2010validity} to the design of this study. Any conclusion made from this work must be considered with the following issues in mind:

\textbf{Conclusion validity} focuses on the significance of the treatment. To enhance conclusion validity, we ran experiments on 10 different target projects and found that our proposed method always performed better than the state-of-the-art approaches. More importantly, we applied a wider step ($d = 0.35$) for the statistical testing of Cohen'd (described in \textbf{RQ1} of \S5) than the state-of-the-art work~\cite{jitterbug} ($d = 0.2$) so the observed effects are validated with more confidence. In addition, we have considered generalization issues of single evaluation metrics (e.g., recall and precision)  and instead evaluate our methods on metrics that aggregate multiple metrics like F1 and G1 while being more effort-aware (APFD and cost). As future work, we plan to test the proposed methods with additional analyses that are endorsed within SE literature (e.g., P-opt20~\cite{tu2020better}) or general ML literature (e.g., MCC~\cite{mcc_metrics}).

Finally, in order to understand why our proposed method outshines the others, we offer a deeper analysis via Table \ref{tab:rq5}. The SOTA active learning Jitterbug starts with the filtering step (Easy) before active learning (Hard) on the rest.  Recall \S2.2.5, one of our hypothesized Jitterbug's shortcomings is Jitterbug's Easy can find up to 90\% of the SATDs can make it difficult for Hard to follow up and find the rest of 10\% SATDs. Hence, Table \ref{tab:rq5}.a reported the percentage of SATDs being found via the filtering step (Easy~\cite{jitterbug} or CLA~\cite{nam2015clami}) and Table \ref{tab:rq4}.b reported the median ratio of estimated SATDs 0over actual SATDs across all iterations per project for Jitterbug's Hard~\cite{jitterbug} (after Easy), \IT(0)'s Falcon, \IT(100)'s Hard (after CLA), and Emblem \cite{tu2020better}. The hypothesis here is that the higher the SATDs being discovered by the early filtering step, the harder it is to find the rest of SATDs, and the higher the overestimation is done by the consequent active learning strategy. This forces the human experts to review more comments than necessary, which results in more cost and effort. For instance, in \textit{JRuby}, Easy finds 90\% of the SATDs so Jitterbug's Hard has a difficult time finding the rest 10\% and ends up overestimating the rest SATDs on average by 8.3 times. Jitterbug's Hard always overestimates (up to 12 times than the actual SATDs) than the rest of the methods except in the case of \textit{Apache Ant} and \textit{EMF}. Meanwhile, CLA only finds up to 29\% of the SATDs on average so the estimation of \IT(100)'s Hard is only 16\% away of the actual SATDs and more balanced (i.e, lowest in IQR). With no prior filtering step, Emblem tends to underestimate by 27\% but \IT(0)'s Falcon has the closest estimation to the actual SATDs , i.e., 95\%. 


\textbf{Internal validity }focuses on how sure we can be that the treatment caused the outcome. To enhance internal validity, we heavily constrained our experiments to the same dataset, with the same settings, except for the treatments being compared.

\textbf{Construct validity }focuses on the relation between the theory behind the experiment and the observation. To enhance construct validity, we compared solutions with and without our strategies in Table~\ref{tab:rq2} and \ref{tab:rq3} while showing that both components (unsupervised learning with \textbf{CLA} and active learning of \textbf{Falcon/Hard}) and in both settings (low-resource labels versus labels abundant) to improve the overall performance. However, we only showed that with our setting of featurization and default parameters of random forest learners. The performance can get even better by tuning the parameters, variety of scalers, and different data preprocessors (e.g., synthetic minority over-sampling or SMOTE that is known to help with unbalanced datasets \cite{agrawal2018better,chawla2002smote}). We plan to explore these in our future work.

\textbf{External validity} concerns how widely our conclusions can be applied. In order to test the generalizability of our approach, we always kept a project as the holdout test set and never used any information from it in training.

\section{Conclusion and Future Work} \label{sec:conclusion}
Managing Self-Admitted Technical Debts are important to maintaining a healthy software project. The current automated solutions do not have satisfactory precision and recall in identifying SATDs to fully automate the process. Moreover, the learning requires the training data to be labeled, which is not always available because of high cost and labor as the case discussed in \S\ref{super_learning}. We showed that there is a ``technical-debt proneness tendency'' in the data where SATDs are associated with higher complexity of the data. In order to reduce the label famine and human effort, a half-automated two-mode framework was proposed, called \IT\hspace{-1pt}. If there is a lack of labeled data, \IT(0) first pseudo-labels the training data's labels using an unsupervised learning method that is based on ``technical-debt proneness tendency''. When there are abundant labeled training data, \IT(100) applies the same unsupervised learner to find the best tendency on the training data to filter out the \textit{highly prone} SATDs from the test data. Then, an active learning model iteratively trains and update on both historically labeled data and new human-labeled ones while guiding the human experts to target the most likely SATDs according to the model's ranking. Our proposed active learning method (i.e, Falcon) is the best one for \IT(0) while Yu et al.'s Hard~\cite{jitterbug} is the best one for \IT(100). The process can be repeated till the estimated recall from the model reaches the predefined target recall. Our findings include:

\be
\item When combining the previous SOTA's pattern-based approach (i.e., Easy) and a simple unsupervised learning (i.e., CLA), the performance was higher than Easy alone without any human effort. Therefore, Easy+CLA should be considered as the new simple baseline method for identifying SATDs. 
\item In case of low-resource data, a combination method of CLA to pseudo-label the training data and a novel active learning strategy (i.e., Falcon) surpassed both the SOTA semi-automated method Jitterbug~\cite{jitterbug} and the SOTA deep learning automated method CNN~\cite{ren2019neural}. This serves as a proof-of-concept of how unsupervised learning can cheaply label the training data to bootstrap the active learning of a machine learning model to identify SATDs on the test data. 
\item In case of having access to the labeled data, a combination method of CLA to filter out the highly prone SATDs and Hard performed similarly to Ren et al.'s CNN~\cite{ren2019neural} while outperforming Yu et al.~\cite{jitterbug} ($5/8$ cheaper). 
\item Falcon is our novel active learning method for its effectiveness in identifying SATDs by frugally pseudo-labeling data and also when having access to the training data labels (as Falcon without filtering performed better than both SOTA methods, in RQ3). 
\item The success of those methods for technical debt suggests that there could be many more domains in software analytics that could benefit from unsupervised learning. As mentioned above, those
benefits include the ability to commission new models, faster, with much less time consuming and less error-prone labeling of examples.
\item Overall, our proposed super learning method with \IT{} is the most effective and efficient in identifying SATDs. 
\ee


That said, {\IT} still suffers from the validity threats discussed in \S\ref{tion:threats}. To further reduce those threats and to move forward with this research, we propose the following future work:
\bi

\item
Apply hyper-parameter tuning on data preprocessing and model configuration to see if our current conclusions still hold and whether tuning can further improve the performance.
\item
Explore more complex patterns (other than just single word patterns \textbf{Easy} has explored).
\item
Survey more advanced feature engineering in the active learning strategy for finding the rest of SATDs. For example, explore N-gram patterns~\cite{wattanakriengkrai2019automatic} and word embeddings with deep neural networks~\cite{flisar2019identification}.
\item Explore other sampling techniques to help with unbalanced class data (one of the key characteristics for SATDs~\cite{ren2019neural}).
\item Test whether replacing the random forest model in {\IT} with a deep learning model (i.e., CNN~\cite{ren2019neural}) will further improve its performance.
\item Try different settings of labeling and filtering (via unsupervised learning) combine with a deep learning model (i.e., CNN~\cite{ren2019neural}) to automatically identify the SATDs. 
\item Survey other unsupervised learning methods for frugally pseudo-labeling and filtering data. 
\item Extend the work to label debt in other artifacts where technical debt may be presented (e.g. issue trackers, documentation) and other software engineering domains (e.g., security, issue close times, static warning analysis, etc) and compare it with other state-of-the-art methods which continue to appear.
\ei

\section*{Acknowledgements}
This work was partially funded by an NSF CISE Grant \#1931425.

\bibliographystyle{plainnat}
\bibliography{main.bib}
 


\clearpage
\appendix

\end{document}